\documentclass[final,nofootinbib,pre,twocolumn,showpacs,showkeys,groupaddress,preprintnumbers,floatfix]{revtex4-1}

\usepackage{dynlearn}

\renewcommand{\H}{\operatorname{H}}

\newcommand{\cs}{\causalstate}

\newcommand{\msym}{\meassymbol}

\newcommand{\Abet}{\ProcessAlphabet}
\newcommand{\SSet}{\CausalStateSet}

\newcommand{\MxSSet}{\AlternateStateSet}
\newcommand{\MxSMeasure}{\mu}
\newcommand{\MxSDyn}{\mathcal{W}}
\newcommand{\mxst}{\eta}

\newcommand{\One}{ {\mathbf{1} } }

\newcommand{\qemach}{cCQS\xspace}
\newcommand{\measuredM}{measured cCQS\xspace}
\newcommand{\measuredMs}{measured cCQSs\xspace}
\newcommand{\MeasuredM}{Measured cCQS\xspace}

\newcommand{\PMProtocol}{\mathcal{E}}
\newcommand{\QSSProcess}{R_{-\infty:\infty}}
\newcommand{\MProcess}{X_{-\infty:\infty}}

\DeclareMathOperator{\Tr}{Tr}
\DeclareMathOperator*{\argmin}{arg\,min} \DeclareMathOperator*{\argmax}{arg\,max} 

\newcommand{\overbar}[1]{\mkern 1.5mu\overline{\mkern-1.5mu#1\mkern-1.5mu}\mkern 1.5mu}

\begin{document}

\def\ourTitle{Optimality and Complexity\\
in\\
Measured Quantum-State Stochastic Processes
}

\def\ourAbstract{\indent If an experimentalist observes a sequence of emitted quantum states via
either projective or positive-operator-valued measurements, the outcomes form a
time series. Individual time series are realizations of a stochastic process
over the measurements' classical outcomes. We recently showed that, in general,
the resulting stochastic process is highly complex in two specific senses: (i) it is inherently unpredictable to varying degrees that depend on measurement
choice and (ii) optimal prediction requires using an infinite number of temporal
features. Here, we identify the mechanism underlying this complicatedness as
generator nonunifilarity---the degeneracy between sequences of generator states
and sequences of measurement outcomes. This makes it possible to quantitatively
explore the influence that measurement choice has on a quantum process' degrees
of randomness and structural complexity using recently introduced methods from
ergodic theory. Progress in this, though, requires quantitative measures of
structure and memory in observed time series. And, success requires accurate
and efficient estimation algorithms that overcome the requirement to explicitly
represent an infinite set of predictive features. We provide these metrics and
associated algorithms, using them to design informationally-optimal
measurements of open quantum dynamical systems.
}

\def\ourKeywords{quantum stochastic process, quantum measurement,
  hidden Markov chain, \texorpdfstring{\eM}{epsilon-machine}, causal states,
  mixed-state presentation, Shannon entropy rate, statistical complexity
}

\hypersetup{
  pdfauthor={James P. Crutchfield},
  pdftitle={\ourTitle},
  pdfsubject={\ourAbstract},
  pdfkeywords={\ourKeywords},
  pdfproducer={},
  pdfcreator={}
}

\title{\ourTitle}

\author{Ariadna Venegas-Li}
\email{avenegasli@ucdavis.edu}

\author{James P. Crutchfield}
\email{chaos@ucdavis.edu}

\affiliation{Complexity Sciences Center and Physics Department,
University of California at Davis, One Shields Avenue, Davis, CA 95616}

\date{\today}
\bibliographystyle{unsrt}

\begin{abstract}
\ourAbstract
\end{abstract}



\preprint{\arxiv{2205.03958}}

\date{\today}
\maketitle

\setstretch{1.1}

\newcommand{\qssp}{quantum-state stochastic process}
\newcommand{\QSSP}{Quantum-State Stochastic Process}
\newcommand{\qssps}{quantum-state stochastic processes}
\newcommand{\QSSPs}{Quantum-State Stochastic Processes}
\newcommand{\QStateSet}{\mathcal{A}_Q}

\section{Introduction}
\label{sec:intro}

Time series of controlled quantum states are essential to quantum physics,
quantum information and computing, and their implementations in novel
technologies. Moreover, as quantum technologies scale to larger qubit
collections that evolve coherently over increasingly longer times,
fault-tolerant design and error correction of quantum-state time series become
increasingly necessary.

Several fault-tolerant systems and diagnostic tools have been developed
assuming quantum processes with uncorrelated noise \cite{Schi11a,
Nigg14a, Ofek16a, Saro20a, Harp20a}. Unfortunately, progressing beyond those
assumptions to more physically realistic non-Markovian, correlated processes
has been challenging. To date, error correction for non-Markovian quantum
processes can be deployed only in specific cases. Moreover, contemporary theory
offers a restricted toolset for quantum process identification and control
\cite{Riva14a, deVe17a, Whit20a}.

The following explores one reason for these challenges and limited progress.
In short, there is substantially more-complicated statistics and
correlational structure embedded in non-Markovian quantum processes and in
the classical stochastic processes that result from measurement than
currently appreciated. We show how to identify the signatures of these
complexities and how to constructively address the challenges they pose.

To ameliorate historical inconsistencies, along the way we give unified
definitions of Markov and non-Markov processes. Said most simply, these address
the role of memory in a structurally consistent way---a way that also gives
access to modern multivariate information theory. Clarity in this is essential
to appreciating the structural varieties of complex quantum processes. Lasting
progress in complex quantum processes depends on this clarity.

Explicating the structures embedded in quantum processes requires stepping
back, to revisit a basic question: How to characterize the stochastic process
that results from measuring sequential quantum states? The answer is found in a
recently-introduced framework for correlated and state-dependent quantum
processes \cite{Vene20a}. Notably, its toolkit relies only on classical
dynamical systems. The turn to the latter is perhaps unexpected from the
perspective of quantum physics, but makes sense given that the goal is to
describe the classical data an experimentalist has in hand in the laboratory.
Specifically, the tools rely on fundamental results that both reach back more
than half a century to early ergodic and stochastic process theories, but also
call on contemporary mathematics of dynamical systems---specifically, iterated
function systems and their stable asymptotic invariant measures
\cite{Jurg20b,Jurg20c,Jurg20e}.

\begin{figure*}[htbp]
\centering
\includegraphics[width=\linewidth]{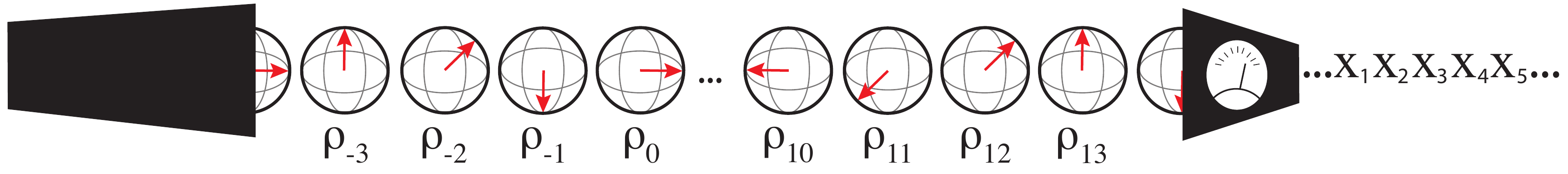}
\caption{A general controlled quantum source (CQS) as a discrete-time quantum
	dynamical system (black box) that stochastically generates a time series of
	quantum states $\rho_{t-2} \rho_{t-1} \rho_t \ldots$ (density matrices).
	Measuring each state in the sequence realizes a classical stochastic
	process over random variables $\ldots X_1 X_2 X_3 X_4 X_5 \ldots$.
	}
\label{fig:QubitBlackBox}
\end{figure*}

Thus, the objects of study here are time series of sequential quantum systems
and the stochastic processes that result from sequentially measuring the state
of each one.

Generating a time series of sequential quantum states usually occurs under the
control of an experimental apparatus. If control over the apparatus is not
perfect or if it undergoes dynamics that are unstable or not fully understood,
then the time series of emitted quantum states can be profitably regarded as a
stochastic process. It may also be desired for a given application, such as
quantum cryptography, that a quantum state process be stochastic.

Figure \ref{fig:QubitBlackBox} illustrates this with a black box quantum system
that emits a quantum state $\rho_t$ (density matrix) at each time $t$. We refer
to this source as a \emph{controlled quantum source} (CQS) and to its output as
a \emph{\qssp} (QSSP). It is important to note here, that a distinct quantum
state is emitted at each timestep and the object of study is the time series of
these states. To emphasize, we are not investigating the dynamical evolution of
an individual quantum state or individual quantum system.

Continuing from left to right in Fig. \ref{fig:QubitBlackBox}, measurement of a
quantum-state time series results in a stochastic process of classical
measurement outcomes. Of that classical process we then ask:
\begin{itemize}
\setlength{\topsep}{0pt}
\setlength{\itemsep}{0pt}
\setlength{\parsep}{0pt}
\item \emph{Statistics}: What are the properties of this observed classical
	process---its randomness, correlation, memory?
\item \emph{System identification}: What properties of the underlying quantum
	stochastic generator hidden in the black box can be reconstructed from the
	observed classical process? 
\item \emph{Measurement choice}: How does measurement affect the observed
	statistics and controller identification?
\end{itemize}

\emph{Computational mechanics} \cite{Crut88a, Shal98a, Crut12a} was originally
introduced to constructively answer these questions for purely-classical
hidden processes. To do this, it extracts the ``effective theory'' from a time
series of observations and provides measures of the time series' randomness and
structure. In particular, a process' \emph{statistical complexity} $\Cmu$
quantifies how much structure or memory is required to do optimal prediction.
And, the \emph{entropy rate} $\hmu$ quantifies a process' intrinsic
randomness---the rate of information production.

This toolset, together with the fact that measuring a quantum time series
results in a classical time series, motivates our approach. We adapt these
classical measures of intrinsic randomness and structure to describe the
classical time series observed when applying measurement operators to a time
series of quantum states. This provides a description of relevant properties of
a stochastic process of quantum states---properties that have proven usefully
diagnostic and descriptive in the classical setting. The following endeavors to
show that they are for quantum processes. Moreover, our approach allows
analyzing the effects that measurement choice has on the observed complexity
of a quantum dynamical process.

This serves as a starting point to more fully appreciating the complicatedness
of quantum dynamical processes and the role that measurement plays. Over the
longer term, building on this, the goal is to characterize stochastic quantum
processes beyond being Markovian or non-Markovian---memoryless or
memoryful---to arrive at understanding of their informational and statistical
properties. These metrics can then support more informed approaches to error
correction and to potentially leveraging noise for particular tasks, 
just as their classical analogs have for
thermodynamic computing \cite{Parr15a, Cont19a}. 

Our agenda is as follows. Section \ref{sec:Background} motivates the set-up and
places the results in the context of the observed stochasticity of quantum
systems. It introduces the notion of quantum stochastic processes and
information and correlations in quantum time series. The next two sections go
into detail on the technical problem statement. First, Sec.
\ref{sec:TempQComplexity} describes the general type of QSSP that we study and
how we model them. Second, Sec. \ref{sec:measurement} details how we measure a
QSSP and describes the classical stochastic process that results. Section
\ref{sec:EmergentNonunifilar} identifies generator nonunifilarity as the
mechanism by which quantum measurement induces varying randomness and
correlation in the resulting classical process. Section \ref{sec:Metrics} then
gives a short summary of the tools required to analyze measured QSSPs.
Following that, Sec. \ref{sec:ProcessClasses} applies the methods to example
QSSPs and their corresponding measured processes. Section \ref{sec:properties},
in particular, highlights the general properties that can be extracted.
Section \ref{sec:Apps}, by way of illustrating the general results, applies
the previously introduced metrics to particular examples and investigates their
dependence on measurement choice and discusses alternative measurement
protocols. This serves as a way to introduce notions of measurement optimality
in Sec. \ref{sec:Optimal}.

\section{Background}
\label{sec:Background}

To start, let's locate our development in contrast to other quantum settings. 

Perhaps the most common setting for quantum processes investigates a single
quantum system evolving in time. In this, stochasticity in the system's state
evolution arises from its interaction with an environment. Within this setting
stochasticity in temporal evolution can also arise from inherent nonlinear
dynamics or repeated measurement and other state mappings.

Historically, though, there is a longstanding effort to characterize
stochasticity in quantum dynamics as a means to manage quantum noise
\cite{Gard04a, Cler10a}. Much of the machinery developed to describe quantum
stochastic phenomena arose from open quantum systems, seeded around quantum
master equations and relying heavily on assuming process Markovity
\cite{Breu07a, Riva12a}. More recently, an effort emerged to understand,
detect, and quantify non-Markovity, and many examples of specific non-Markovian
quantum dynamics have been analyzed in detail \cite{Riva14a, Breu16a, deVe17a,
Li18a}.

From the perspective of Markov's original concept of ``complex chains''
\cite{Mark13a,Mark13b} and the modern theory of (classical) Markov processes,
though, these approaches rely on an unnecessarily-varied set of Markovity
definitions. One consequence is that, most basically, they do not agree on what
process memory and correlation are. This also makes comparing results across
investigations challenging. And so, we give and apply a unified definition
of memory that addresses classical and measured quantum processes. Appendix
\ref{app:Markovity} discusses this in more detail, noting the conflicting
definitions one finds.

Directly importing the concept of classical stochastic processes to the quantum
domain has proved challenging, as Ref. \cite{Milz20a} discusses in detail.
While there are several causes, the primary difficulty is that the Kolmogorov
Extension Theorem---specifying event-sequence probabilities and
measures---breaks down when the random variables are quantal and are generated
by quantum mechanisms.

Recently, process tensors were introduced to treat such quantum stochastic
processes \cite{Poll18a, Poll18b, Tara19a}. They promise a gateway to probe
quantum stochasticity beyond the binary distinction of Markov versus
non-Markov. Modeling correlations beyond merely Markovian will allow analyzing
a broader class of quantum processes. That said, the endeavor is new.

While there are parallels with process-tensor representations, the following
focuses on a different type of dynamical process---different also from
individual quantum-system temporal evolution. It considers a sequence of
distinct entities, in which the quantum state of each is a random variable at
each time step. In this, there is a rough parallel to quantum spin chains.
Notably, these variables can be correlated by the physical mechanism that
sequentially generates them.

Consider a physical system that emits qubits, each in a state that is noisy or
stochastic; for instance, photons emanating from a blinking quantum dot
\cite{Efro16a}. The difference between these processes and the more-familiar
evolving single-system dynamics is that, at each time step, the quantum state
of the emitted qubit is its own random variable. That is, each successive
random variable lives in its own Hilbert space. The associated random variable
takes on a specific qubit state, and measuring these states does not interfere
with the quantum process generator nor with other prior or future qubits.
In short, as the quantum source emits quantum states the dimension of
the product Hilbert space grows. As a consequence, addressing time-asymptotic
properties requires working with an infinite product state space. And, these
change the kind of investigation we pursue. We investigate time series of
quantum states emitted by a physical system in the spirit of analyzing the
multivariate statistical and informational properties of the system's dynamics
as developed in Ref. \cite{Crut01a}.

Specifically, Ref. \cite{Vene20a} recently introduced a framework for such
quantum-state stochastic processes. It casts the problem of quantum processes
in a way that is amenable to directly applying the tools of classical
stochastic processes to characterize the informational and structural
properties of QSSPs. It showed that a qubit time series, when observed through
projective measurements, generically results in a highly complex classical
stochastic process. Highly complex here means that the observed process has
positive entropy rate and requires an infinite number of temporal features to
optimally predict future outcomes. It also demonstrated that measurement
choice---the manner in which an observer interacts with a qubit stream---can
drive a quantum process appear more or less complex.

To accomplish this, the prequel adapted the metrics of computational mechanics
to describe randomness and structure in the measured processes. Section
\ref{sec:Metrics} summarizes these metrics. The following takes those results
further by introducing an analytical framework to describe stochastic processes
over quantum states. It provides a more thorough-going exploration and isolates
the physical mechanism---generator nonunifilarity---responsible for the
observed complexity. It analyzes a variety of examples that span the possible
types of dynamics and offers several avenues for future explorations, including
outlining how to optimize measurements of quantum processes to various ends.

\section{Quantum-State Stochastic Processes}
\label{sec:TempQComplexity}

The following introduces the main objects of study---\qssps---and the
information sources that generate them. It then moves on to explain how the
classical processes that emerge ``in the lab'' are produced through measurement
of individual quantum states in QSSP realizations. With the latter processes in
hand, it then shows how to quantify their randomness and structure using metrics
available from information theory and explores how those properties vary as a
function of measurement choice.

\subsection{Quantum Processes}
\label{ssec:qprocess}

Consider a given quantum source that emits a sequence of individual quantum
states. At each time step, the quantum state it emits takes on a value from a
finite set. We refer to these sources as \emph{controlled quantum sources}
(CQSs) and, in their operation, they generate \emph{\qssps}. We will now define
\qssps.

Let $R_t$ denote the random variable for the quantum state emitted at time $t$.
The realization of $R_t$ as a particular quantum state is $\rho_t \in
\QStateSet$, where $\QStateSet$ is the set of available quantum states in a
Hilbert space. The random variable for a sequence of quantum states emitted
between times $t$ and $t+\ell$ is denoted by the block random variable
$R_{t:t+\ell}$ (inclusive on the left, exclusive on the right). Then
$\rho_{t:t+\ell}$ denotes the realized sequence of quantum states.

\begin{Def}[Homogeneous \QSSP]
\label{def:qsp}
Let $\mathcal{A}_Q \subseteq \mathcal{H}^d$ be the set of available quantum
states in the $d$-dimensional Hilbert space $\mathcal{H}^d$. $\Omega =
\mathcal{A}_Q^{\mathbb{Z}}$ is then the space of bi-infinite sequences over
$\mathcal{A}_Q$. Consider the probability space $\mathcal{P} = (\Omega,
\mathcal{F}, P)$, where $\mathcal{F}$ is the $\sigma-$algebra on the cylinder
sets of $\Omega$ and $P$ a probability measure over the cylinder sets.
$\QSSProcess$ denotes the discrete-time random-variable sequence
of quantum states described by the \qssp\ $\mathcal{P}$. It comprises the
sequences of random variables that take on values according to a measurable
function $T_t : \Omega \rightarrow \mathcal{A}^t_Q$:
\begin{align*}
R_t = T_t(\QSSProcess)
  ~,
\end{align*}
for $t \in \mathbb{Z}$.
\end{Def}

To emphasize again, we work with distinct physical objects at each time step.
That is, each random variable $R_t$ takes a value on its own Hilbert space
$\mathcal{H}^d_t$. In a \emph{homogeneous} process the Hilbert spaces are of
the same type. Think, for example, of a time series of photons, each photon
emitted by a source at a certain time step, in this or that quantum state. 

We restrict our study to \emph{stationary} and \emph{ergodic} QSSPs. 
 
\begin{Def}[Stationarity]
\label{def:stationary}
A stationary QSSP is one in which the probability of observing a particular
sequence of quantum states is independent of the time at which the observation
is made. That is, the probability of an observed quantum sequence is
time-translation invariant:
\begin{align}
\Pr (R_{t:t+\ell} = \rho_{t:t+\ell}) = \Pr (R_{0:\ell} = \rho_{0:\ell})
\label{eqn:stationary}
  ~,
\end{align}
for all $t\in\mathbb{Z}$, $\ell \in \mathbb{Z}$, and $\rho_{0:\ell}$.
\end{Def}

\begin{Def}[Ergodicity]
\label{def:ergodic}
A QSSP is \emph{ergodic} if all long realizations obey the QSSP's statistical
properties. That is, given a long realization $R_{0:n} = \rho_{0:n}$, the
probability of observing a finite realization of $R_{0:\ell}$ of length $\ell
\ll n$ in $\rho_{0:n}$ is the same as observing that same block in multiple
realizations of length $\ell$ drawn from the QSSP.
\end{Def}

The following considers only QSSPs satisfying these three definitions. It also
imposes two more restrictions. First, it focuses on stochastic processes of
qubits, since they are the basic carriers of quantum information. Second, it
focuses on processes in which for all $t$: $\rho_t^2 = \rho_t$. That is, the
quantum states realized by the $R_t$s are strictly pure states, lying on the
surface of the Bloch sphere. This rules out (for now) the presence of
entanglement within quantum state blocks. And, in practical terms, it means the
joint variable over qubits emitted in a time interval can be represented as:
\begin{align}
R_{t:t+\ell} = R_t \otimes R_{t+1} \otimes \ldots \otimes R_{t+\ell-1}
\label{eqn:separable_states}
\end{align}
and a realization is both a pure state and the tensor product of the individual pure quantum states:
\begin{align*}
\rho_{t:t+\ell} = \rho_{t} \otimes \rho_{t+1} \otimes \ldots
  \otimes \rho_{t+\ell-1}
  ~.
\end{align*}

Again, together these definitions describe a setup in which a quantum system
emits individual qubits at each time step. The latter are in pure quantum
states. The quantum system that emits the qubits is the controlled qubit source
(CQS). Occasionally, this is abbreviated as the \emph{controller} to make
reference to an experimenter having some level of control, perhaps through
design, over qubit emission, which can be stochastic.

\subsection{Quantum Sources and Generator Presentations}
\label{ssec:ModelQProcess}

\begin{figure*}[htbp]
\centering
\includegraphics[width=\linewidth]{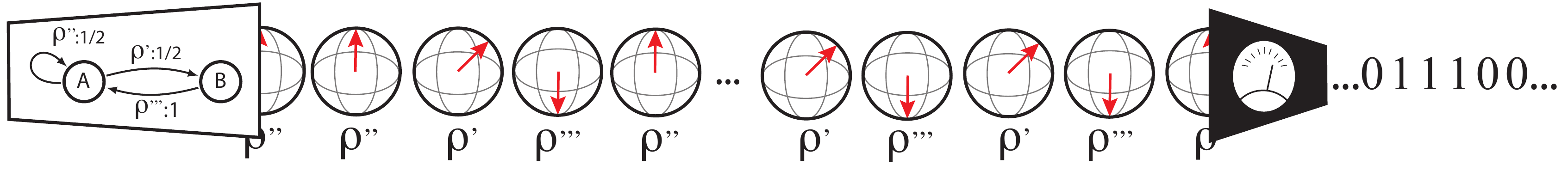}
\caption{Classically-Controlled Quantum Source and its Emitted Process: (Left)
	As the internal hidden Markov controller operates, at each time $t$ the
	emitted symbols $\mathcal{A}$ (not shown) determine the qubit's quantum
	state $\rho_t \in \mathcal{A}_Q$. As shown, the
	latter are realized as density matrices $\rho{'}$, $\rho{''}$, and
	$\rho{''}$. (For simplicity the controller's emitted symbols $\mathcal{A}$
	are not explicitly shown, only the resulting quantum states $\rho \in
	\mathcal{A}_Q$.) (Middle) The resulting output quantum process is a
	sequence of distinct $\mathcal{H}^2$ Hilbert spaces, displayed as a series
	of Bloch spheres with realized state vectors displayed inside. (Right)
	Measuring each qubit realizes a classical stochastic process $\ldots
	011100$.
	}
\label{fig:QubitFiniteStateController}
\end{figure*}

In short, quantum sources generate quantum processes. Here, we concentrate on a
particular implementation of CQSs---those that generate QSSPs via a classical
controller with a finite memory in the form of a hidden Markov chain (HMC)
\cite{Rabi86a,Rabi89a,Bech15a}. (Appendix \ref{app:hmm} provides a short
refresher on HMCs.) To illustrate, this means that the black box in Fig.
\ref{fig:QubitBlackBox} is reified as in the white box shown in Fig.
\ref{fig:QubitFiniteStateController}: a HMC controlling what the box emits.
Correspondingly, we refer to this qubit source as a \emph{classically
controlled qubit source} (\qemach).

The choice of finite-state HMC controller to model a \qemach is natural, in
that HMCs are a standard representation of finite-memory stochastic processes,
are widely used to model noisy classical sources and communication channels,
and can be fully analyzed \cite{Riec13a,Riec17a}. Also, once we complete
appropriately developing the present framework, HMCs are easily extended to
represent more general sources, such as those driven by quantum controllers
\cite{Moor97a}.

\begin{Def}[Classically-Controlled Quantum Source]
\label{def:ccqs}
A cCQS is a tuple $( \CausalStateSet, \mathcal{A}_Q, \{\mathbb{T}^\rho\})$
where:
\begin{enumerate}
\setlength{\topsep}{0pt}
\setlength{\itemsep}{0pt}
\setlength{\parsep}{0pt}
\item $\CausalStateSet$ is the set of hidden states of the HMC controller.
\item Alphabet $\mathcal{A}$ is a finite set of symbols in one-to-one
	correspondence with the set $\mathcal{A}_Q$ of available qubit quantum
	states. The latter are emitted by the cCQS when a symbol is encountered on
	a transition. To simplify notation, both a symbol and its qubit state are
	denoted by a density matrix $\rho \in \mathcal{H}^2$. 
\item $\{\mathbb{T}^\rho: \rho \in \mathcal{A}_Q\}$ is a set of quantum-state
	labeled transition matrices of size $|\CausalStateSet| \times
	|\CausalStateSet|$. $\mathbb{T}^\rho_{\cs\cs'}$ is the probability of
	transitioning from internal state $\cs$ to internal state $\cs'$ (both in
	$\CausalStateSet$) while emitting symbol (or, effectively, quantum state)
	$\rho$.
\end{enumerate}
\end{Def}

The labeled transition matrices $\{\mathbb{T}^{\rho}\}$ sum to the
internal-state stochastic transition matrix over hidden states: $\mathbb{T} =
\sum_{\rho \in \mathcal{A}_Q} \mathbb{T}^{\rho}$. This, in turn, determines the
HMC's stationary internal-state distribution $\pi$ as the left eigenvector of
$\mathbb{T}$ with eigenvalue $1$: $\pi = \pi \mathbb{T}$. $\pi$ is then a
vector of size $|\CausalStateSet|$ in which the entry $\pi_\cs$ represents the
asymptotic probability of the HMC being in internal state $\cs \in
\CausalStateSet$.

\begin{figure}
  \centering
  \includegraphics{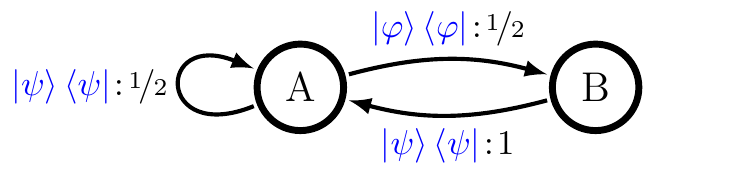}
\caption{cCQS Operation: If the HMC controller is in state $A$ the cCQS
	has equal probabilities of remaining there or transitioning to state
	$B$. If a transition to state $B$ then occurs, the system emits a qubit in
	the pure state $\ket{\varphi}\bra{\varphi}$. In the next time step the
	cCQS must transition to state $A$ and it then emits a qubit in state
	$\ket{\psi}\bra{\psi}$. That is, the edge labels $\rho:p$ indicate taking
	the state-to-state transition with probability $p$ and emitting quantum
	state $\rho$. As the cCQS operates, a qubit is output at each time step,
	over time the result is a qubit time series.
	}
\label{fig:qem0}
\end{figure}

For an example, see the cCQS in Fig. \ref{fig:qem0}, where:
\begin{enumerate}
\setlength{\topsep}{0pt}
\setlength{\itemsep}{0pt}
\setlength{\parsep}{0pt}
\item $\CausalStateSet = \{A, B\}$.
\item $\mathcal{A}_Q = \{\rho_\psi = \ket{\psi}\bra{\psi},
	\rho_\varphi = \ket{\varphi}\bra{\varphi}\}$. Here, $\ket{\psi}$ and
	$\ket{\varphi}$ are pure quantum states.
\item $\{\mathbb{T}^{\rho}\} = \{\mathbb{T}^{\rho_\varphi},
\mathbb{T}^{\rho_\psi}\}$, where:
\begin{align*}
\mathbb{T}^{\rho_\varphi} & = \begin{pmatrix} 0 & 1/2 \\ 0 & 0 \end{pmatrix} \\
\mathbb{T}^{\rho_\psi} & = \begin{pmatrix} 1/2 & 0 \\ 0 & 1 \end{pmatrix}
  ~.
\end{align*}
\item $\pi = \begin{pmatrix} 2/3 & 1/3 \end{pmatrix}$, the left eigenvector of:
\begin{align*}
	\mathbb{T} & = \mathbb{T}^{\rho_\varphi} + \mathbb{T}^{\rho_\psi} \\
	& = \begin{pmatrix} 1/2 & 1/2 \\ 0 & 1 \end{pmatrix}
	~.
\end{align*}
\end{enumerate}

The following implements cCQSs with finite-memory HMC controllers:
$|\CausalStateSet| < \infty$. It also specifies that HMC controller transitions
are \emph{unifilar}: The current internal hidden state and emitted quantum
state uniquely determine the next internal hidden state.

Section \ref{sec:Unifilarity} gives a more general and detailed account of
unifilarity. Section \ref{sec:Metrics} then highlights several of its
consequences. As will become clear, the distinction between unifilar and
nonunifilar HMCs plays a large role in driving the complexity of quantum
processes on their own and when measured.

These choices ensure that the \qemach's randomness and complexity can be
directly calculated. And so, the effects of measurement on the quantum process
are made most explicit. That said, the analysis below can be applied directly
to nonunifilar controllers, with the caveat that the controller's structure and 
randomness are more complicated to quantify as will be made clear in the following sections. 

\section{Measured Quantum-State Stochastic Processes}
\label{sec:measurement}

A central aspect of the process realized in the laboratory is how an observer
interacts with the QSSP. Naturally, interactions occur via quantum measurement,
but there are multiple options for both measurement type and implemented
protocol. We now set the stage for measurement protocols and define the
specific measurements and protocols used.

\subsection{Measurement Protocols}

We view a measurement protocol as a communication channel between the quantum
stochastic process $\QSSProcess$ and its measured companion---a classical
stochastic process $\MProcess$. Denote the relationship between the QSSP random
variables and those of the measured process by: 
\begin{align}
X_t = \mathcal{I}_t(R_t)
  ~.
\label{eqn:measured_qsp}
\end{align}

Where $\mathcal{I}_t$ represents the action of a measurement on the quantum
state $\rho_t$ output at time $t$. The set $\mathcal{I} = \{\mathcal{I}_t\}$
for all $t$ defines a measurement protocol. In a slight abuse of notation, we
use the variable $\mathcal{I}_t$ to represent both the measurement channel and
the particular measurement operator applied at time $t$. For short, we refer to
the measurement protocol as just $\mathcal{I}$ and denote the relationship
between the QSSP and its corresponding measured stochastic process by: 
\begin{align}
\MProcess = \mathcal{I}(\QSSProcess)
  ~.
\label{eqn:channel}
\end{align}

To recapitulate the notation, at each time step the random variable $R_t$ takes
on the value of a particular quantum state $\rho_t$. That, in turn, is measured
with operator $\mathcal{I}_t$. The resulting measurement outcome is denoted by
the random variable $X_t$, which takes on a particular value $x_t \in
\mathcal{A}_M$, with $\mathcal{A}_M$ the set of possible measurement outcomes. 

We now define the basic measurement protocol used.

{\Def[Single State Constant-Measurement Protocol] As a QSSP is output, each
quantum state passes through a measurement channel $\mathcal{I}_t =
\mathcal{E}$, for all $t$. That is, the same measurement  $\mathcal{E}$
is applied to each output state individually: $X_t = \mathcal{E}(R_t)$.
}

Though the following employs only this protocol, we note that it is
straightforward to work with measurement protocols for which $\mathcal{I}_t$
depends on time. For example, a measurement scheme in which the measurements
alternated between a measurement along the $z$-axis and a measurement along
the $x$-axis. Or, a potentially more useful protocol is one in which
measurements are adaptive, and the $\mathcal{I}_t$s are chosen given the
outputs of a subset of past measurements. One approach to the adaptive
measurement protocols is described in Ref. \cite{Gier21a}.

\subsection{Projective Measurements}
\label{ssec:measurement}

At each time-step, the observer performs a single measurement $\mathcal{E}$ on
the quantum state $\rho$ emitted by the controller. This measurement consists of
a finite set of nonnegative operators $\{E_x\}$, in which the index $x \in
\mathcal{A}_M$ labels the measurement outcomes. The measurement operators sum
to the identity: $\sum_{x \in \mathcal{A}_M} E_x = \mathbb{I}$. The probability
of measurement outcome $x$ after measuring quantum state $\rho$ is given by
$\Pr(x|\rho) = \Tr(E_x\rho)$, where $\Tr(\cdot)$ is the trace.

We concentrate our analysis on single-state projective measurements, in which
the operators are orthogonal. This simplifies the basic framework, for now.
That said, Section \ref{sec:povms} briefly considers single-state protocols with
more-general positive operator-valued measurements (POVMs).

{\Def[Projective Measurement on a QSSP]

A \emph{projective measurement} $\mathcal{E}$ in a dimension $d$ Hilbert space
$\mathcal{H}^d$ consists of a set of mutually orthogonal projectors $\{E_x\}$,
with $x \in \{0, 1, \ldots d-1\}$, such that  $E_x E_y = E_x\delta_{xy}$ and
$\sum_x E_x = \mathbb{I}_d$. When the measurement $\mathcal{E}$ acts on a
quantum system in state $\rho_t$, emitted by the QSSP, the outcome is $x_t$
with probability  $\Pr(x_t|\rho_t) = \Tr(E_{x_t}\rho_t)$. Applying a projective
measurement to every state emitted by the QSSP yields a classical stochastic
process over the values of $x$. We call the set of possible values of $x$ the
\emph{measured process alphabet} $\mathcal{A}_M$.
}

For example, consider a projective measurement of a qubit consisting of two
orthogonal measurement operators $\{E_0, E_1\}$. Without loss of generality
these can be written as: 
\begin{subequations}
\begin{align*}
E_0 & = \ket{\psi_0} \bra{\psi_0} \\
E_1 & = \ket{\psi_1} \bra{\psi_1} 
  ~.
\end{align*}
\end{subequations}
Later, we refer to the set $\mathbf{E} = \{E_i: i = 0, 1, 2, \ldots\}$
of such measurement operators as the \emph{observation basis}.

Let $\ket{\psi_i} \in \mathcal{H}^2$ and $\braket{\psi_0 | \psi_1} =
0$.  When working with qubit projective measurements, it is convenient to
parametrize them using Bloch angles as follows: 
\begin{subequations}
\begin{align}
\ket{\psi_0} & = \cos{\frac{\theta}{2}}\ket{0}
	+ e^{i\phi}\sin{\frac{\theta}{2}} \ket{1}\\
\ket{\psi_1} & =  \sin{\frac{\theta}{2}}\ket{0}
	- e^{i\phi}\cos{\frac{\theta}{2}} \ket{1}
  ~.
\end{align}
\label{eqn:param}
\end{subequations}

The resulting measurements are labeled $0$ or $1$, respectively. Let $X_t$
denote the random variable for the outcome of measuring the state $\rho_t$ at
time $t$ and $x_t$ the realized value. In the case of qubit projective
measurements $x_t \in \{0,1\}$. In this way, applying projective measurement
protocol to a QSSP produces a binary classical stochastic process. Knowledge of
the CQS's controller and the measurement protocol are the basic ingredients
needed to analyze the mechanism that generates these classical stochastic
processes---the measured processes.

\subsection{Measured Processes}
\label{ssec:measured}

By Eq. (\ref{eqn:channel}) the classical stochastic process that is realized by
measuring the QSSP is the joint distribution $\Pr(\MProcess)$. The specific
value $x_t$ taken on by the random variable $X_t$ depends both on the
measurement protocol $\mathcal{I}=\{\mathcal{I}_t\}$ with measurement operators $\{E_{x_t}\}$ and
the QSSP and its HMC controller. That is, if the random variable $R_t$ takes on
the particular value $\rho_t$ at time $t$, then:
\begin{align}
\Pr(X_t=x|R_t = \rho_t, \mathcal{I}_t) = \Tr(E_{x_t} \rho_t)
  ~.
\end{align}

Both the measurement protocol and the QSSP can introduce correlations within
the classical stochastic process. That is, even if applying a time-independent
measurement protocol, such as a constant single-state measurement protocol, the
correlations in $\QSSProcess$ will yield correlations in $\MProcess$. However,
even if $\QSSProcess$ is an independent, identically distributed (IID) process,
a time-correlated measurement protocol $\mathcal{I}$ can yield a correlated
classical stochastic process.

\subsection{Measured Process Presentations}
\label{ssec:ModelsMeasured}

Importantly, in cases where the QSSP is generated by a cCQS, the measured
quantum process can be modeled with a unique HMC, as the following demonstrates.

{\Prop{Applying a projective measurement protocol $\PMProtocol$ to a QSSP
$\QSSProcess$ generated by a finite-state cCQS $M$ results in a measured process
$\MProcess$ given by a finite-state HMC. 
}}

{\ProProp{We establish this by directly constructing the HMC presentation. The
HMC $\mathbb{M} = \{ \CausalStateSet, \mathcal{A}_M, \{T^x\}, \pi \}$ that
generates the measured process has the following components:
\begin{enumerate}
\setlength{\topsep}{0pt}
\setlength{\itemsep}{0pt}
\setlength{\parsep}{0pt}
\item The same set $\CausalStateSet$ of internal states as the HMC that
	generated the QSSP.
\item A finite alphabet $\mathcal{A}$ consisting of each possible measurement
	outcome.
\item A set of labeled transition matrices $\{T^x\} $with $x \in \mathcal{A}_M$
	such that: 
\begin{align}
    T^x = \sum\limits_{\rho \in \mathcal{A}_Q} \mathbb{T}^{\rho} \Prob(x|\rho, \PMProtocol)
    \label{eqn:trans_probs}
\end{align}
with: 
\begin{align}
	\Prob(x|\rho, \PMProtocol) = \Tr(E_x \rho)
	~.
\end{align}
\item The same stationary distribution $\pi$ as the HMC that generated
	the QSSP.
\end{enumerate}
}}

{\Def{We refer to the resulting HMC as a \emph{measured cCQS}.
}}

\section{Emergent Quantum Complexity}
\label{sec:EmergentNonunifilar}

In this way, fixing a cCQS and a measurement basis determines a unique measured
cCQS. This HMC accurately describes the classical stochastic process resulting
in the lab.

One would hope to directly analyze the statistical properties of the classical
process using that HMC. Or, more modestly, to better and more accurately analyze
the classical process using the HMC than by simulating repeated realizations
over long times to obtain statistics for estimation. The HMC, after all,
exactly describes the process, being a presentation.

We demonstrate that this analysis is very far from a straightforward procedure.
Moreover, the difficulties are (i) inherent and (ii) generic to quantum
measurement. Despite the challenges, though, with care and the right tools in
hand one can fully characterize the measured process' properties.

We introduce two classes of HMCs---those that are unifilar (already
peripherally introduced above) and those that are not. The following then
explains why measurement induces complex statistics. Specifically, the
following establishes that (i) nonunifilarity arises in the measured process
HMC, (ii) this is generic for projective measurements, and (iii) complex
statistics arise in the measured process due to an exponential explosion of
the predictive feature set. Along the way, we introduce generative and
predictive presentations---those that can be used to produce process
realizations and those that, in addition, can be used to optimally predict
realizations.

\subsection{Presentations}
\label{sec:Presentations}

A given stochastic process can be generated by many different HMCs. Each is
called a \emph{presentation} of the given process. The essential requirement is
that a presentation describes all and only a process' realizations and their
probabilities. HMC presentations are either \emph{unifilar} or
\emph{nonunifilar}. Unifilarity controls how useful a presentation is to
quantitatively analyzing a process.

\subsection{Presentation Unifilarity}
\label{sec:Unifilarity}

{\Def[Unifilarity] An HMC transition is unifilar if the current internal state
$\cs$ and the emitted symbol $x$ uniquely determine the next internal state
$\cs'$: $\Pr(\cs'|x,\cs) = 1$, for at most one $x$. That is, there is at most
one transition leaving a state for each symbol.

An HMC is \emph{unifilar} if all its transitions are.}

When an HMC is \emph{nonunifilar} there is ambiguity in the next state for at
least one transition.

It is a notable fact---one motivating the distinction in the first place---that
processes generated by finite unifilar HMCs are less (typically much less)
complex than those that can be generated only by finite nonunifilar HMCs.

An intuitive way to see why this occurs is the following. Consider a
realization of a given process. If it was emitted by a unifilar HMC the
realization has a one-to-one or at most one-to-finite correspondence between
the observed symbol series and sequences of hidden-state transitions. In
contrast, a realization generated by a nonunifilar HMC has a one-to-infinite
correspondence between observed symbols and hidden state transitions. The
result in this case is that the number of possible of sequences of hidden
states that emit a particular sequence of observed symbols grows exponentially
with sequence length. In general, a significantly more complex hidden structure
is required to optimally predict processes generated by nonunifilar HMCs.

\subsection{Predictors, Generators, and Irreducible Nonunifilarity}
\label{sec:Unifilar}

A stochastic process' unifilar presentation, up to redundancies in states
or transitions, is an optimal predictor. That is, given an HMC hidden
state, the probabilities of the next observed symbols are the optimal, most
informed prediction of what that next observed symbol will be.

Unifilar HMCs being process predictors contrasts with nonunifilar HMCs which
are not predictors. The latter are only generators of process realizations.
Moreover, their states are typically poor predictors.

One way to restate the distinction between process predictors (unifilar
presentations) and process generators (nonunifilar presentations) is the
following. On the one hand, for a unifilar presentation, there is a
deterministic relation between the past $x_{-\infty:t}$ and the current hidden
state $\cs_t$. That is, $\cs_t = f(x_{-\infty:t})$, where $f(\cdot)$ is a
function; while many pasts $x_{-\infty:t}$ may lead to $\cs_t$.  Moreover, for
all such pasts, $\cs_t$ must have the same conditional distribution
$\Prob(X_{t:\infty}|\cs_t) = \Prob(X_{t:\infty}|x_{-\infty:t})$ of future
sequences given the observed past. Since we can use $\Prob(X_{t:t+1}|\cs_t)$
to predict future observations, we say that the hidden states in a
unifilar presentation are \emph{predictive}. 

On the other hand, when employing a process' nonunifilar presentation to
predict its future $\Prob(X_{t:t+1}|\cdot)$ requires a mixture of distributions
$\Prob(X_{t:t+1}|\cs_t)$ over the presentation's states $\{\cs_t\}$. In this
sense, nonunifilar states are not predictive. Nonunifilar presentations still
\emph{generate} the process accurately, since the states and transitions
explicitly provide a probabilistic procedure for eventually emitting all of a
process' realizations with the correct probabilities.

Starting with any unifilar HMC presentation for a stochastic process, one can
eliminate redundancies in information about the past by merging states $\cs_t$
with identical future probability distributions $\Prob(X_{t:\infty}|\cs_t)$.
Eliminating these redundancies gives a unique minimal optimal predictive HMC
for a stochastic process. This canonical presentation is called a process'
\emph{\eM} \cite{Crut88a, Crut12a}. The \eM's states are a process'
\emph{causal states}. They are causal is the sense that they give the minimal
mechanism that allows one to exactly predict process realizations.

Looking ahead, we must distinguish between processes with finite predictive
presentations and those without. And, this we can now do.

{\Def{A stochastic process is \emph{irreducibly nonunifilar} if it is
generated by a finite nonunfilar HMC, but there exists no finite unifilar HMC
presentation that predicts it.}}

\subsection{Measurement-induced Nonunifilarity}
\label{sec:Nonunifilar}

With this background, the tools are in place to address the origins of
measurement-induced complexity in observed quantum-state stochastic processes.
First, we identify the emergence of nonunifilarity. Second, we argue this
happens frequently and, in fact, is generic in measured QSSPs. Third,
we explore the consequence---explosive complexity. Finally, we identify the
underlying mechanism driving this in quantum state indistinguishability.

{\Prop{\label{prop:EmergentNonunifilarity} Quantum measurement of a QSSP
generated by a cCQS can lead to a classical process represented by a
nonunifilar measured cCQS.}}

{\ProProp{Consider cCQS hidden state $\cs$ with outgoing transitions to two
distinct hidden states $\cs_I$ and $\cs_{II}$. The first transition emits
quantum state $\rho'$ and the second, $\rho''$. Now, performing a measurement
$\{E_0, E_1\}$ on the emitted quantum states, both measured transitions have
nonzero probability of emitting the same symbol. Recall from Eq.
(\ref{eqn:trans_probs}):
\begin{align*}
T^x_{\cs \cs'} = \sum\limits_{\rho \in \mathcal{A}_Q}
  \mathbb{T}^{\rho}_{\cs \cs'} \Prob(x|\rho)
  ~.
\end{align*}

Note that above and in what follows we suppress explicit mention of the measurement 
protocol in the conditional probabilities, simplifying $\Prob(x|\rho,\PMProtocol)$ to 
$\Prob(x|\rho)$ for ease of notation. 
Now, consider a $\rho$ that gives a nonzero probability of obtaining
measurement outcome $x$. Note that this is the case for $\cs' = \cs_I$ and
$\cs' = \cs_{II}$. Then, as long as $\mathbb{T}^{\rho}_{\cs \cs'}$ is nonzero
for this $\rho$, both $T^x_{\cs \cs_I}$ and $T^x_{\cs \cs_{II}}$ will be
nonzero. This makes the observed transition out of state $\cs$ nonunifilar and,
thus, makes the measured cCQS nonunifilar. 
}}

The implications become more apparent later, when discussing how HMC
nonunifilarity almost always implies that the process it generates is
irreducibly nonunifilar. In this way, measurement can---and as discussed later,
almost always will---induce irreducible nonunifilarity of the measured
process. 

\subsection{Nonunifilarity is Generic}
\label{sec:Generic}

We say a property is \emph{measurement generic} over a set of
measurements---for instance, qubit projective measurements---if it holds true
for almost all measurements but a measure zero subset. Similarly, a property is
\emph{source generic} if it holds true for the QSSPs generated by almost all
cCQSs.

{\Prop{\label{prop:genericity}
A measured cCQS HMC is generically nonunifilar.
This is true both source generically and measurement generically over the set
of projective measurements.
}}

{\ProProp{Consider the constraints that give rise to unifilar transitions.
Recall the entries of the labeled-transition matrices for a measured
cCQS, as defined in Eq. (\ref{eqn:trans_probs}):
\begin{align*}
T^x_{\cs \cs'} = \sum\limits_{\rho \in \mathcal{A}_Q}
  \mathbb{T}^{\rho}_{\cs \cs'} \Prob(x|\rho)
  ~.
\end{align*}
Each entry is composed of a sum of terms. For the measured cCQS to maintain unifilarity there should be at most one nonzero term per row of each labeled transition matrix. That is, for each $x \in \Abet$ and $\cs \in \CausalStateSet$ pair, the term $T^x_{\cs \cs'}$ is nonzero for at most one value of $\cs'$.

For unifilarity to hold, the following conditions on the underlying cCQS and
measurement must be satisfied:
\begin{enumerate}
	\setlength{\topsep}{0pt}
	\setlength{\itemsep}{0pt}
	\setlength{\parsep}{0pt}
\item \label{itm:first} A hidden state with an outgoing transition to only one
	other hidden state maintains unifilarity.
\item \label{itm:second} All cCQS hidden states can have at most two outgoing
	transitions (to distinct states). Denote the quantum states emitted on the
	outgoing edges $\rho_a$ and $\rho_b$ and the two destination states $\cs_a$
	and $\cs_b$, respectively. Thus, for each cCQS hidden state $\cs$ there are
	at most two nonzero transition elements: $\mathbb{T}^{\rho_a}_{\cs \cs_a}$
	and  $\mathbb{T}^{\rho_b}_{\cs \cs_b}$, say. When determining the measured
	cCQS's labeled transitions $T^x_{\cs \cs'}$, for each $\cs$ and $x$, there
	will be at most two contributions. The following condition ensures that
	these two contributions do not result in two or more nonzero values for
	each $\cs$.
\item \label{itm:third} If the state has two outgoing transitions then, to
	maintain unifilarity, it must satisfy: 
	\begin{itemize}
		\setlength{\topsep}{0pt}
		\setlength{\itemsep}{0pt}
		\setlength{\parsep}{0pt}
	\item The two emitted quantum states $\rho_a$ and $\rho_b$ on the
		outgoing transitions must be orthogonal to each other.
	\item The measurement basis must be aligned with $\rho_a$ and $\rho_b$. 
		That is, each measurement operator must project into the quantum states
		that the process emits---$\rho_a$ and $\rho_b$.
	\end{itemize}
	For a given pair of $x$ and $\cs$, these ensure that the only potentially
	nonzero terms are $T^{\rho_a}_{\cs \cs_a}$ and $T^{\rho_b}_{\cs \cs_b}$.
	If the projective measurement with outcome $x$ is aligned with either
	$\rho_a$ or $\rho_b$, however, then only one of the terms of the form
	$\mathbb{T}^{\rho}_{\cs \cs'} \Prob(x|\rho)$ can be nonzero for a given
	$\rho$. And so, there will be at most one nonzero term in the $\cs$ row of
	transition matrix $T^x$. This guarantees that the measured cCQS remains
	unifilar.
\end{enumerate}

Conditions \ref{itm:first}, \ref{itm:second}, and \ref{itm:third} are highly
restrictive in the space of cCQSs. That is, almost none of the possible labeled
transition matrices $\mathbb{T}^{\rho}$ satisfy them. In turn, this means that
measured cCQSs are source-generically nonunifilar. 

For a given cCQS, Condition \ref{itm:third} is highly restrictive in the set of
projective measurements and is only satisfied for one measurement choice out of
a continuous set of possible measurement choices. Therefore, the measured cCQS
is measurement-generically nonunifilar. 

It is also important here to note that Conditions \ref{itm:second} and
\ref{itm:third} refer to the case in which the cCQS emits qubits. For a cCQS
that emits qudits, these conditions can be generalized to allow for $d$
distinct transitions to other states. The generalization is straightforward: To
maintain unifilarity the output quantum states associated with those $d$
transitions must be mutually orthogonal and the measurement chosen must be able
to distinguish perfectly between those $d$ states.
}}

As Sec. \ref{sec:Explosive} develops in more detail, Prop.
\ref{prop:genericity} says that measured processes are typically highly
complex, in the sense that they generically have an uncountable infinity of
predictive features (causal states), divergent statistical complexity, and a
positive entropy rate. 

\subsection{Variations}
\label{sec:Variations}

Several observations are in order on generic nonunifilarity for qubit processes
and how generic nonunifilarity trades-off against the Hilbert space dimension
of the QSSP's quantum states.

Structurally, a binary alphabet highly restricts the possibilities for a
particular HMC's topology to support unifilarity. This could lead to a rushed
conclusion that restricting to projective measurements plays a determinant role
in nonunifilarity of measured quantum processes. In fact, however, allowing for
POVMs does not change this aspect. Suppose a particular \qemach hidden state
has two outgoing edges with nonorthogonal quantum states $\rho_a$ and $\rho_b$.
In any POVM with two or more measurement operators at least one has a nonzero
probability of being an outcome when applied to both $\rho_a$ and $\rho_b$.
This means that in the measured \qemach there are at least two distinct outgoing
transitions with the same symbol. And, this again yields nonunifilar dynamics.
Section \ref{sec:povms} explores this for an example POVM measurement protocol.

In general, when the quantum states emitted by the \qemach are restricted to
qubits, the relatively low dimensionality of the Hilbert space means that we
generically recover nonunifilar machines. This is due to the fact that the only
case in which a measurement with two or more operators (not necessarily
projectors) can perfectly distinguish between two quantum states is when they
are orthogonal. (Distinguishing quantum states here means that none of the
operators have nonzero probability to be the measured outcome on both of the
quantum states.) And, even in this case, distinguishability holds only for a
particular measurement basis that aligns with the two orthogonal states to
measure.

The second observation concerns a potentially-useful generalization of how to
reduce nonunifilarities in higher dimension. The preceding establishes that
irreducible nonunifilarity dominates in measured QSSPs, and this is true
generically. However, if the quantum states emitted by the \qemach are qudits,
there is more ``room'' to reduce the nonunifilar transitions in the \measuredMs
when the Hilbert space dimension is larger than $d=2$. For instance, when the
number of outgoing transitions from one hidden state to distinct hidden states
is at most $d$, one can partition the set of quantum states emitted in those
transitions into mutually orthogonal subsets. In that case one can devise a
measurement that captures each orthogonal subset as a distinct measurement
outcome. This, in a sense, constrains the nonunifilarities to be only in the
outgoing transitions that have output quantum states with nonzero overlap.
Effectively, the \measuredM loses information about which specific state was
output within each orthogonal subset by turning those into nonunifilar
transitions, but maintains the information about which output states where
mutually orthogonal. This somewhat reduces the complexity of the \measuredM.

As a simple example, consider a \qemach that outputs qutrits. A particular
hidden state has three outgoing edges to distinct states, each outputting
qutrits in states $\ket{0}$, $\ket{+}$, and $\ket{2}$. One can devise a
measurement that outputs $0$ if the qutrit is in the subspace spanned by
$\ket{0}$ and $\ket{1}$, and outputs $1$ if the qutrit is orthogonal to that
subspace. In this case the \measuredM has nonunifilarity only in the $\ket{0}$
and $\ket{+}$ transitions. This reduction of nonunfilarity is only viable as
long as there is an orthogonal subset within the set of possible output states
$\mathcal{A}_Q$. So, it is still rare, but the larger the Hilbert space of
output states is, the more opportunities there are for reducing nonunifilarity.

Naturally, these measurements should also be physically motivated by the
information the experimenter is trying to extract from the underlying quantum
dynamic. Thus, when working with quantum information from a classical reality,
there is a tradeoff between the complexity of the observed dynamics and how
coarsely or finely one probes the quantum state through measurement. Compared
to general qudits, the space of qubits offers much less room for coarser
measurements. 

The practical upshot of these arguments is that analyzing a measured \qemach
requires working with nonunifilar presentations of the observed classical
stochastic process.

\subsection{Explosive Complexity}
\label{sec:Explosive}

Consider a process that is generated by a finite nonunifilar presentation.  One
can construct a unifilar presentation for it. The reasons for doing so will
become abundantly clear shortly. For now, say a predictive presentation is
needed. The states of the unifilar presentation are Blackwell's \emph{mixed
states} \cite{Blac57b}. These are identified using the \emph{Mixed State
Algorithm} introduced in Refs. \cite{Crut08a, Crut08b} and explained in detail
in Ref. \cite{Jurg20b}. Said simply, by tracking the ``states of knowledge''
about an HMC's internal states as revealed indirectly by emitted symbols, one
builds a unifilar hidden Markov chain whose states are the mixed states and
whose transitions are the mixed-state to mixed-state transitions. The result is
known as the process' \emph{mixed state presentation} (MSP). The MSP then
provides an insightful and calculationally efficient way to determine many, if
not all, of a process' statistical and informational properties.

Appendix \ref{app:msa} reviews MSP construction and its properties, here we
summarize. Given a process' $N$-state HMC presentation $M$, one constructs
$M$'s set $\MxSSet$ of mixed states as the conditional probability
distributions $\eta(x_{-\ell:0}) = \Pr(\CausalState_0 = \cs |X_{-\ell:0} =
x_{-\ell:0})$ over the HMC's hidden states $\cs \in \CausalStateSet$ given all
possible sequences $x_{-\ell:0} \in \mathcal{A}^\ell$. Given $M$ and an
observed symbol sequence $x_{-\ell:0}$, there is a unique mixed state
$\eta(x_{-\ell:0})$ that represents the best guess as to $M$'s current internal
state. Moreover, the set of the process' allowed sequences of all lengths $\ell
\in \mathbb{N}$ induces a invariant measure $\mu$ on the state distribution
$(N-1)$-dimensional simplex. We simply denote this as the \emph{mixed state
distribution} $\mu(\MxSSet)$. An HMC's mixed state set $\MxSSet$ together with
the transition dynamic $\mathcal{W}$ between mixed states induced by observed
sequences form the HMC's MSP: $\mathrm{MSP}(M) = \{\MxSSet,\mathcal{W}\}$. 

Importantly, by construction an HMC's MSP is a unifilar presentation of the
stochastic process generated by the HMC. Additionally, the set of mixed states
$\MxSSet$ corresponds to the process' set of causal states. The consequence is
that the MSP, up to minimizing state redundancies, is the unique optimally
predictive model of the stochastic process---its \eM \cite{Jurg20c}.

{\Conj{A process generated by a nonunifilar presentation generically is
an irreducibly nonunifilar stochastic process. That is, it requires an infinite
number of predictive features (causal states) for optimal prediction.}
}

Blackwell introduced this conjecture in his seminal 1957 work on classical
stochastic processes \cite{Blac57b}. There, he developed several of the first
information-theoretic results for what he called \emph{functions of Markov
chains}. These are equivalent to what are nowadays called hidden Markov chains.
Moreover, for very specific cases Blackwell showed that the set of (predictive)
features that a process stores from observed sequences can be finite or
countable. In all other instances, the predictive features set is uncountably
infinite. These predictive features are equivalent to the process' MSP mixed
states $\MxSSet$. The primary lesson is that the predictive complexity of
irreducibly nonunifilar processes explodes, despite them being generated by a
finite mechanism---a finite-state HMC.

Long experience and extensive explorations of HMC space support Blackwell's
claims and this conjecture, which has also been recorded elsewhere
\cite{Crut92c, Marz17a, Vene20a, Jurg20b, Jurg20c}. Reference \cite{Jurg20e}
goes into great detail about the mechanisms by which these stochastic processes
generate and process information. It reviews the arguments and evidence that
the conjecture holds quite broadly. Finally, for \measuredMs we have not
encountered a single violation. That said, establishing the conjecture for the
general or the quantum settings remain open problems.

\subsection{Quantum State Indistinguishability}
\label{sec:Indistinguishability}

Section \ref{sec:Generic} detailed the structural reasons that make measured
cCQSs generically nonunifilar. Behind these lies a simple physical property
that is responsible for irreducible nonunifilarity and, thus, explosion in
complexity of measured quantum processes. When applying a measurement to a
QSSP that emits qubits in two or more distinct quantum states, a single
measurement will generally have a nonzero probability of not being able to
distinguish which quantum state it measured. This indistinguishability between
quantum states therefore acts as a source of noise. And, this makes direct
reading of the QSSP's underlying structure markedly more memory intensive.
This, in turn, radically increases the predictive complexity of the measured
process with respect to the QSSP. 

One can quantify how distinguishable or indistinguishable two quantum states
are using the trace distance \cite{Niel11a} for instance. If a particular
hidden state in a cCQS has outgoing transitions to two distinct hidden states
that emit two different quantum states, a measurement makes the distinction
ambiguous (noisy) unless the trace distance between the two quantum states is
unity. In that case, the quantum states have orthogonal supports. Moreover,
there is the further requirement for not inducing nonunifilarity that the
measurement distinguish between the two states. If these criteria are met,
then nonunifilarity in the measured process is not created and there is no
explosion in predictive complexity. However, these criteria are very
restrictive and so explosive complexity is to be expected in \measuredMs.

\section{Measured Quantum Process Randomness and Structure}
\label{sec:Metrics}

Simply establishing explosive complexity is insufficient. One needs yardsticks
for analysis and comparison. This section introduces metrics for quantifying
randomness and structure in the classical stochastic processes resulting from
measured QSSPs. The mathematics for these metrics depend critically on the
stochastic process presentation, whether it is unifilar or nonunifilar. The
latter is particularly relevant, as the above showed that the measured
processes are overwhelmingly irreducible nonunifilar. We begin introducing
entropy rate and statistical complexity and how to compute them from unifilar
HMCs. The bulk of the effort and interest, though, arise in adapting these to
nonunifilar presentations, which follows shortly.

\subsection{Unifilar Generators}

When an HMC is unifilar, there is a one-to-one or one-to-finite correspondence
between a sequence of observed symbols and the sequence of hidden states that 
generated it. This allows direct, closed-form calculation of process intrinsic
randomness and predictive memory from the HMC's internal Markov chain.

\subsubsection{Information Creation}
\label{ssec:UniInfoCreation}

Process randomness---the rate at which the process generates information---is
quantified through the Shannon \emph{entropy rate} $\hmu$. It is defined
directly for a process, but the useful goal is to obtain short
cuts---expressions in terms of a presentation's states and transitions.

{\Def{
A process' \emph{entropy rate} is \cite{Shan48a}: 
\begin{align}
\hmu = \lim\limits_{\ell \rightarrow \infty} \frac{\H[X_{0:\ell}]}{\ell}
  ~,
\label{eqn:hmu_def}
\end{align}
where $\H[X] = - \sum_{x} \Pr(X = x) \log_2 \Pr(X = x)$
is the Shannon block entropy \cite{Crut01a}.
}}

That is, $\hmu$ is the average uncertainty per observed symbol. Or, said
differently, it quantifies how much information an observer gains
asymptotically with each newly measured symbol.

Reference \cite{Jurg20b} explores in detail how to compute the entropy rate
of a process generated by a given HMC. Here, we summarize. 

For unifilar HMCs, Shannon \cite{Shan48a} showed the entropy rate is exactly
computable in closed-form from the HMC's transition matrices and stationary
state distribution $\pi$:
\begin{align}
\hmu = - \sum_{\sigma \in \CausalStateSet}
  \pi_\sigma \sum_{x \in \mathcal{A}}
  \sum_{\sigma^\prime \in \CausalStateSet}
  T^{(x)}_{\sigma \sigma^\prime} \log T^{(x)}_{\sigma \sigma^\prime}
  ~.
\label{eqn:uni_hmu}
\end{align}
This is the state-averaged transition uncertainty. The stationary state
distribution is determined by the left eigenvector (associated with eigenvalue
$1$ and normalized in probability) of the internal state transition matrix $T
= \Sigma_{x \in \mathcal{A}} T^{(x)}$.

\subsubsection{Information Storage}
\label{ssec:InfoStorage_uni}

To quantify the structure of a process' presentation $M$, the most
straightforward measure is its number $|\CausalStateSet|$ of hidden states.
Beyond that, a more insightful metric is the amount of historical memory or
information the presentation states contain. This is given by the Shannon
entropy of the state distribution:
\begin{align}
\H[\CausalStateSet] = -\sum_{\sigma \in \CausalStateSet}
  \pi_\sigma \log_2 \pi_\sigma
  ~.
  \label{eq:ent_states}
\end{align}
It quantifies how much information the hidden states store about past
observations. That is, it measures how much memory a given HMC has. And, since
unifilar presentations are predictors, $\H[\CausalStateSet]$ is an upper bound
on the amount of information one must maintain on average to optimally predict
the process. This upper bound will typically overestimate the memory of the 
process unless $M$ is a minimal optimal predictive presentation.

For these metrics to describe actual properties of the stochastic process in
question and not those of a particular HMC presentation---that, say, could
have an overly-large and redundant set of states---we use a process' \eM,
reviewed in App. \ref{app:CMech}. In brief, a process' \eM is its minimal
optimal predictive HMC \cite{Crut88a, Shal98a, Crut12a}.

The \eM's hidden states are a process' \emph{causal states} since they
optimally capture the process' causal structure. With them one can make the
most accurate predictions of future symbols and their associated probabilities
while using minimal memory.
The memory in the causal states is then the minimal amount of information from
the past that must be stored in hidden states to optimally predict the future. 

{\Def{A stochastic process' \emph{statistical complexity} $\Cmu$ is the Shannon
entropy of its \eM's causal states $\CausalStateSet$: 
\begin{align}
\Cmu & = \H[\CausalStateSet] \nonumber \\
  & = -\sum_{\sigma \in \CausalStateSet}
  \Prob(S = \sigma) \log_2 \Prob(S=\sigma)
  ~.
\label{eqn:cmu}
\end{align}
}}

$\Cmu$ is the minimal memory required to optimally predict the future. 

\subsection{Nonunifilar Generators}
\label{ssec:nuni_generators}

If the only description available for a measured QSSP is a nonunifilar HMC
presentation, though, then quantifying the process' stochasticity and
structure becomes markedly more complicated due to the explosive complexity
demonstrated above. And, this has substantial practical consequences. In the
case of intrinsic randomness, Eq. (\ref{eqn:uni_hmu}) overestimates the
entropy rate. In the case of structure, the Shannon entropy of the nonunifilar
HMC's hidden states only quantifies the memory used by that particular (likely
nonunique) presentation. More to the point, it does not provide information on
how much memory is minimally required to \emph{optimally} predict the process.

And, if these challenges were not enough, there is yet another complication at
this stage. While constructing the MSP from a process' nonunifilar HMC produces
a unifilar HMC, it is rarely finite-state. As we showed, the typical case is an
infinite-state HMC, generically with uncountably infinite states \cite{Crut92c,
Marz17a, Jurg20b}. Generally then, the \eM, the minimized MSP, has an
uncountable set of causal states. As a consequence, the statistical complexity
of Eq. (\ref{eqn:cmu}) diverges and the expression for entropy rate in Eq.
(\ref{eqn:uni_hmu}) becomes inadequate. 

\subsubsection{Information Creation}
\label{ssec:NuniInfoCreation}

Reference \cite{Jurg20b} showed that the correct expression for the process'
entropy rate is an integral of the transition uncertainty over the
mixed-state simplex $\MxSSet$ weighted by the invariant measure $\mu(\eta)$:
\begin{align}
\hmu^B = - \int_{\MxSSet} d \mu(\mxst) \sum_{\msym \in \MeasAlphabet}
  \Pr(\msym |\mxst) \log_2 \Pr(\msym |\mxst)
  ~.
\label{eqn:hmuB_def}
\end{align}
(The $B$ superscript here is a nod to Blackwell's contribution.)

As introduced in Ref. \cite{Jurg20b}, general contractivity of the MSP dynamic
$\mathcal{W}$ on the simplex and ergodicity allow accurately evaluating the
integral expression. This is implemented by taking an average over a time
series of mixed states $\eta_t$, rather than integrating over the Blackwell
measure $\mu(\eta)$. This yields the process' entropy rate:
\begin{align}
\widehat{\hmu^B} = - \lim_{\ell \to \infty}
	\frac{1}{\ell}
	\sum_{\msym \in \MeasAlphabet} \sum_{i = 0}^\ell
  \Pr(\msym |\mxst_i) \log_2 \Pr(\msym |\mxst_i)
  ~,
\label{eqn:hmuB_eval}
\end{align}
where $\Pr(\msym |\mxst_i) = \mxst(\msym_{0:i})\cdot T^{(\msym)}\cdot \bf{1}$,
$\msym_{0:i}$ represents the first $i$ symbols of an arbitrarily long sequence
$\msym_{0:\ell}$ generated by the process' MSP, and $\bf{1}$ is a column vector
of all ones.

\subsubsection{Information Storage}
\label{ssec:InfoStorage_nuni}

To quantify the structure and memory in these infinite-state processes, not
all is lost due to $\Cmu$'s divergence. While the latter is generally the
case, we can quantify the divergence rate with the \emph{statistical
complexity dimension} $d_\mu$ of the Blackwell measure $\mu(\eta)$ on
$\MxSSet$ \cite{Jurg20c}: 
\begin{align}
d_\mu = -\lim_{\epsilon \to 0} \frac{H_\epsilon[\MxSSet]}{\log_2{\epsilon}}
  ~.
\label{eqn:dmu_def}
\end{align}

This tracks the rate at which the memory requirements for optimal prediction
grow with increasing precision $-\ln \epsilon$. Specifically, $H_\epsilon[Q]$
is the Shannon entropy of the continuous-valued random variable $Q$
coarse-grained at size $\epsilon$. Evaluating $d_\mu$ is not a simple matter,
though. The procedure is presented in detail in Refs. \cite{Jurg20c, Jurg20e}.
In particular, Ref. \cite{Jurg20e} introduces the \emph{ambiguity rate}, which
quantifies the rate at which optimal predictive models discard information by
introducing uncertainty over the infinite past. The difference between the entropy
rate and the ambiguity rate in a process is fundamental to determine its
statistical complexity dimension, as well as the cardinality of its set of mixed states $\MxSSet$. 
The following sections use these methods, suitably adapted to the present
quantum setting.

To give a firmer, even visual, grounding to the preceding results and metrics,
the next section explores three examples representative of distinct classes of
\measuredMs and how the above metrics characterize them.

\section{Classifying Measured Quantum Processes}
\label{sec:ProcessClasses}

The metrics for randomness and structure of a measured quantum process depend
on the cardinality of the mixed state set $\MxSSet$ generated by the
\measuredM. There are three distinct classes: processes for which the number
of mixed states is finite, countably infinite, and uncountably infinite. The
following examples illustrate processes in these classes.

\begin{figure}[htbp]
\centering
\includegraphics{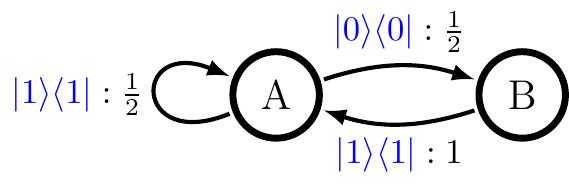}
\caption{Unifilar presentation for the \emph{Observation Basis Golden Mean
	(OB-Golden Mean) process}: A simple \qemach.
	}
\label{fig:ex11}
\end{figure}

\subsection{Finite-State}
\label{ssec:ex1}

\begin{figure}[htbp]
\centering
\includegraphics{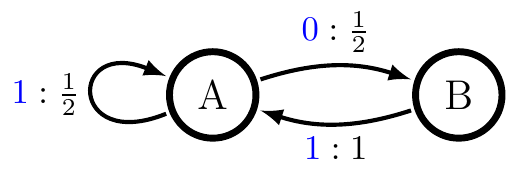}
\caption{Measured \qemach\ of the stochastic process resulting from measuring
	the quantum process generated in Fig. \ref{fig:ex11} in the observation
	basis.
	}
\label{fig:ex12}
\end{figure}

The first quantum process is generated by the unifilar \qemach\ shown in Fig.
\ref{fig:ex11}. It consists of all random sequences without consecutive
$\ket{0}\bra{0}$s. Measuring in the observation basis $E_0 = \ket{0}\bra{0}$
and $E_1 = \ket{1}\bra{1}$) yields a unifilar HMC that generates the
\emph{Golden Mean Process} consisting of all random sequences without
consecutive $0$s. Figure \ref{fig:ex12} shows its minimal presentation---its
\eM: a unifilar HMC with two states. Being unifilar one readily calculates
that it has an entropy rate of $\hmu = 2/3$ bits/symbol from Eq.
(\ref{eqn:uni_hmu}) and a statistical complexity of $\Cmu = 0.918$ bits from
Eq. (\ref{eqn:cmu}).

Although unnecessary in this case, computing the MSP of this presentation---or
any other finite unifilar HMC, for that matter---results in an HMC with a
finite number of states. In the present case both the measured process' entropy
rate and the statistical complexity are finite. They are readily computed via
Eqs. (\ref{eqn:uni_hmu}) and (\ref{eqn:cmu}), respectively.

Section \ref{sec:Generic} showed that quantum processes in this class are
relatively rare in the space of \measuredMs. They occur only under very
constrained circumstances. This observation will become clearer as we consider
more complex classes.

\subsection{Countably-Infinite-State}
\label{subsec:ex2}

\begin{figure}[htbp]
\centering
\includegraphics{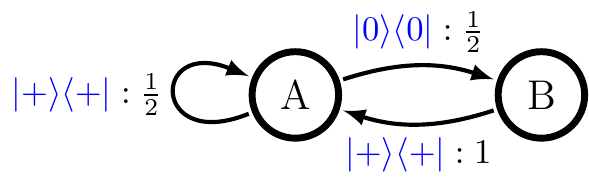}
\caption{Structurally, this \qemach is similar to that in Fig. \ref{fig:ex11}.
	However, not all emitted quantum states are orthogonal. This guarantees
	that the measured process is more complex, as Fig. \ref{fig:ex22} shows.
	}
\label{fig:ex21}
\end{figure}

\begin{figure}[htbp]
\centering
\includegraphics{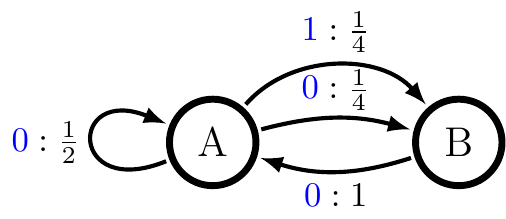}
\caption{Measured \qemach for the process generated by measuring the quantum
	process generated by the \qemach in Fig. \ref{fig:ex21}. This HMC is
	nonunifilar: if in state $A$ and emitting a $0$, the next
	hidden state may be $A$ or $B$.
	}
\label{fig:ex22}
\end{figure}

The next quantum process is generated by the \qemach in Fig. \ref{fig:ex21}.
This is a seemingly slight variation on the previous example. Now, the quantum
alphabet $\mathcal{A}_Q$ consists of nonorthogonal states. Instead of emitting
quantum states in the observation basis, this \qemach emits qubits in state
$\ket{0}\bra{0}$ and others in state $\ket{+}\bra{+}$. In this, we define
$\ket{+}$ and $\ket{-}$ in the conventional way: $\ket{\pm} =
(1/\sqrt{2})(\ket{0} \pm \ket{1})$. When the process generated by this
\qemach\ is measured in the basis $E_0 = \ket{+}\bra{+}$ and $E_1 =
\ket{-}\bra{-}$, the \measuredM\ has the HMC presentation shown in Fig.
\ref{fig:ex22}.

\begin{figure}[htbp]
\centering
\includegraphics[scale=0.9]{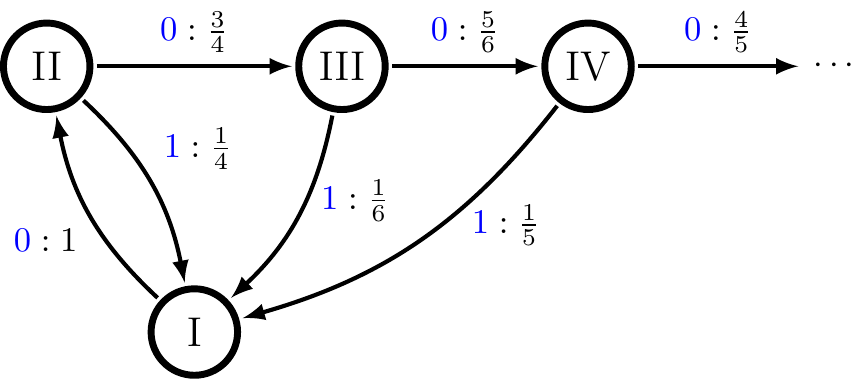} 
\caption{Mixed state presentation of the process generated by Fig.
	\ref{fig:ex22}'s \measuredM.
	}
\label{fig:ex23}
\end{figure}

Notice, though, that Fig. \ref{fig:ex22}'s \measuredM is nonunifilar.
Specifically, knowledge of being in state $A$ and emitting a $0$ does not
determine the next HMC state. The next state could be either $A$ or $B$. Thus,
to compute the entropy rate for this process one must construct its MSP. The
latter is shown in Fig. \ref{fig:ex23}. It has a countable infinity of causal
states. Helpfully, as annotated there, the state transition probabilities can
be parametrized analytically. 

Using Fig. \ref{fig:ex22} one can follow the logic for constructing the MSP.
Independent of any knowledge of the HMC state, seeing symbol $1$ the observer
concludes with absolute certainty that the \measuredM is in state $B$. This is
what we referred to previously as a state of knowledge (or a mixed state)
represented by hidden state I in Fig. \ref{fig:ex23}. In point of fact, the
mixed state associated with state I is $\eta(1) = (0,1)$. After that,
observing symbol $0$ or a sequence of $0$s means that the \measuredM has a
certain probability of being in each \qemach state $A$ or $B$. Each additional
observation of a $0$ then updates the present state of knowledge to one of the
mixed states II $= \eta(10)$, III $= \eta(100)$, IV $= \eta(1000)$, $\ldots$
depending on how many $0$s are observed before seeing a $1$, when the MSP
resets to state I $=\eta(100 \ldots 01)$.

The measured process' entropy rate can be computed from the HMC in Fig.
\ref{fig:ex22} using the methods for nonunifilar HMCs described in Sec.
\ref{ssec:nuni_generators}. Note, though, that for processes whose MSP has a
countable infinity of states, as here, a more rudimentary, though convergent
and accurate, approach is available.

\begin{figure}[htbp]
\centering
\includegraphics[scale=0.5]{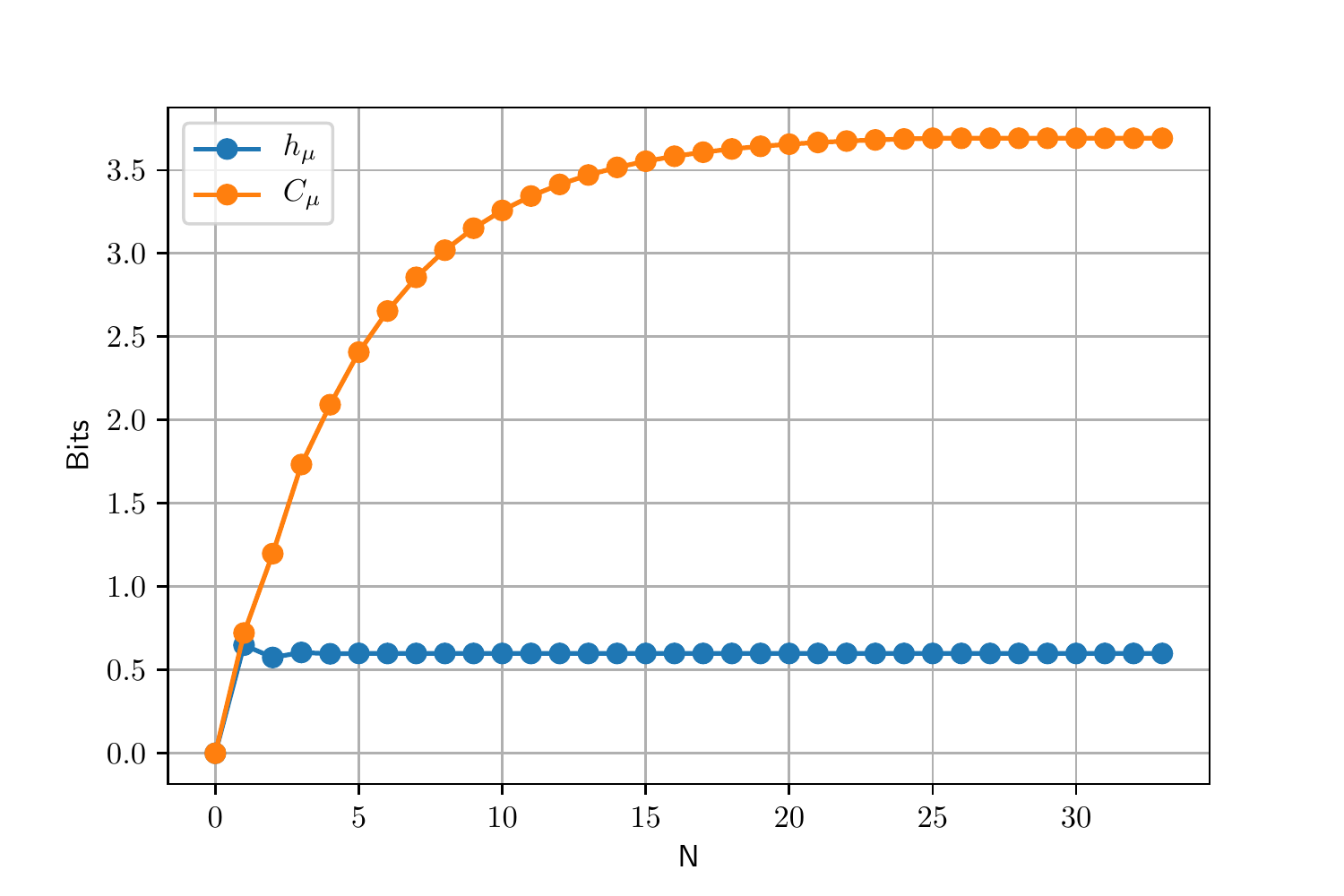} 
\caption{Entropy rate $\hmu$ (blue) and statistical complexity $\Cmu$ (orange)
	of $N$-state HMC approximations of the MSP shown in Fig. \ref{fig:ex23}.
	Notice that $\hmu$ converges rapidly, while $\Cmu$ has a stronger
	dependence on the number of states, but stabilizes around $N = 25$. The
	values obtained are $\hmu = 0.599$ bits/symbol and $\Cmu = 3.69$ bits.
	}
\label{fig:ex24}
\end{figure}

When observing the stochastic process, the probability of observing
consecutive $0$s diminishes with the length of the observed sequence. One then
approximates the process' HMC by truncating the MSP at a finite number $N$ of
mixed states and then exploring the limiting behavior of both $\hmu$ and
$\Cmu$ from those unifilar machines as $N \to \infty$. For the example in
question, this analysis is illustrated in Fig. \ref{fig:ex24}. One finds that
$\hmu = 0.599$ bits/symbol and $\Cmu = 3.69$ bits. Note that, although
infinite state, the process statistical complexity is finite. This is due to
the fact that the asymptotic state distribution $\pi$ decays exponentially
fast for mixed states reached via increasingly more $0$s.

\subsection{Uncountably-Infinite-State}
\label{ssec:ex3}

The preceding two processes are relatively simple, in that they all exhibit a
finite or countable set of mixed states. In the typical case, as argued in Sec.
\ref{sec:EmergentNonunifilar}, the \measuredM has an HMC presentation that is
nonunifilar and an MSP with an uncountable infinity of states. Section
\ref{sec:Generic} established that this is the typical case for processes
generated by cCQSs of two or more states.

\begin{figure}[htbp]
\centering
\includegraphics[scale=0.92]{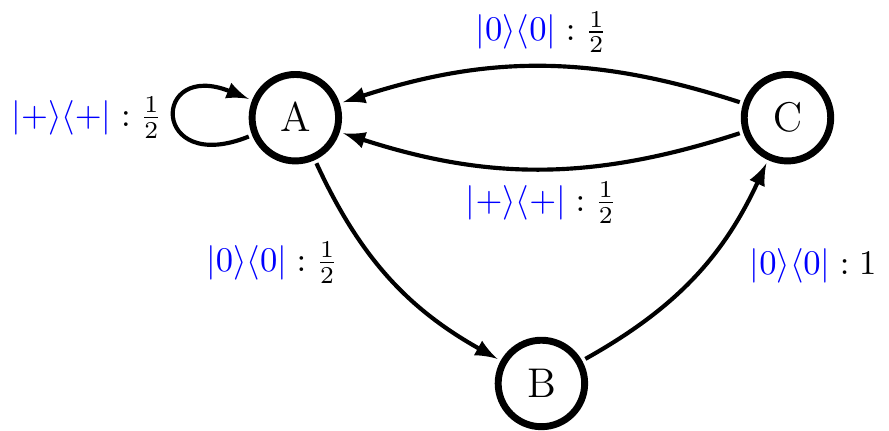} 
\caption{Nonorthogonal Nemo Quantum Process: Three-state \qemach that emits
	qubits in states $\ket{0}\bra{0}$ and $\ket{+}\bra{+}$.
	}
\label{fig:ex31}
\end{figure}

\begin{figure}[htbp]
\centering
\includegraphics[scale=0.92]{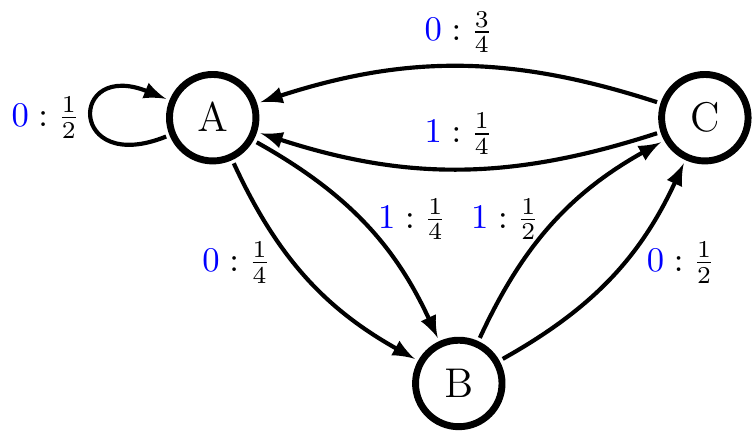} 
\caption{\MeasuredM presentation of the stochastic process produced when
	measuring the quantum process generated by Fig. \ref{fig:ex31}'s \qemach
	measured in the observation basis.
	}
\label{fig:ex32}
\end{figure}

To illustrate, consider the \qemach of Fig. \ref{fig:ex31}, chosen to have
three states principally to aid visualizing the MSP's complexity. The \qemach
is then measured in the observation basis, which yields the \measuredM of
Fig. \ref{fig:ex32}.

\begin{figure}[htbp]
\centering
\includegraphics[scale=0.35]{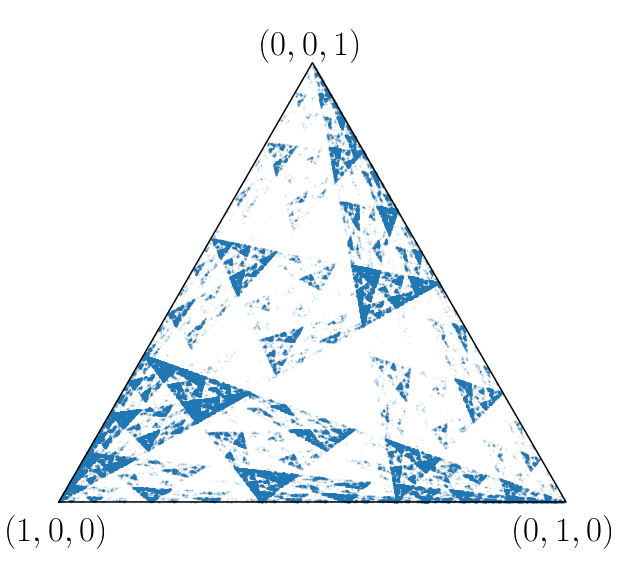} 
\caption{MSP's asymptotic invariant measure $\mu(\eta)$ in the mixed-state
	simplex $\mu(\eta)$. Each mixed state is a point of the form $(p_A, p_B,
	p_C)$ with $p_\cs$ the probability of being in state $\cs$ of the
	\measuredM in Fig. \ref{fig:ex32}.
	}
\label{fig:ex33}
\end{figure}

Note that, as in the example of the countably-infinite state process, the
\measuredM has only a single source of nonunifilarity: the successor state is
ambiguous when observing symbol $0$ with the HMC in state $A$. More generally,
however, none of the symbols $0$ or $1$ allow the observer to ``synchronize''
to the process. That is, observation of a particular symbol does not give an
observer certainty in the \measuredM's state. As argued above mathematically
and as is now constructively clear in Fig. \ref{fig:ex33}, this effectively
translates into the fact that the MSP of the measured quantum process has an
uncountable infinity of mixed states. The MSP---these states together with
their transition probabilities---are a markedly less tractable presentation
than in the previous two quantum processes.

The MSP with all of its states and state transitions cannot be explicitly
displayed as with the previous HMCs. Nonetheless, Fig. \ref{fig:ex33} gives a
sense of the MSP's structure and complexity. It presents a plot of $2 \times
10^6$ MSP states in the mixed-state simplex $\MxSSet$. In fact, it shows $\mu(\eta)$ and its variation in probability density via a histogram with a coarse-graining of $1000 \times 1000$ bins. 

The measured process' entropy rate is computed using Eq. (\ref{eqn:hmuB_eval})
and has a value of $\hmu = 0.8896$ bits/symbol. The statistical complexity
dimension $d_\mu$ of Eq. (\ref{eqn:dmu_def}) is computed as described in Refs.
\cite{Jurg20c,Jurg20e}: $d_\mu = 1.38$.

\subsection{Remarks}

As seen from the three examples above, the cardinality of the mixed state set
of the distinct measured stochastic processes can vary from a finite state set
to a countable infinity of states and on to an uncountable infinity of states.
This cardinality affects the way in which the metrics of randomness and
structure for the process are computed, but also the values they can take.

For processes with finite sets, the statistical complexity and the entropy
rate will generally be finite positive values. This implies that these
processes have a certain degree of stochasticity, but that they can be
optimally predicted with finite memory resources.

Notably, excepting very special cases, this is also true for processes whose
mixed state set has a countably-infinite number of states, as in the second
example. These processes have positive entropy rate, signaling that they have
an intrinsic degree of randomness. And, while they do require an infinite
number of causal states to optimally predict, these states are structured such
that one can simulate an optimal predictor of arbitrary precision with a finite
amount of memory. 

The third case is significantly more complicated than the previous two. A
mixed-state presentation with an uncountable infinity of states implies that the
statistical complexity of the process diverges. This means that it takes
infinite memory to optimally predict these processes. That said, there is an
asymptotic invariant measure over the mixed states in $\MxSSet$. And, by being
able to compute these measures, one can then estimate the process' entropy rate 
$\hmu$ and also the growth rate $d_\mu$ of the memory required for optimal
prediction.

This third case, of processes with MSPs that have an uncountably infinite
number of states, turns out to be the norm for MSPs of measured cCQSs, as
argued above and as we elaborate shortly below. The implications of this are
that, in general, the classical processes that we recover from measuring QSSPs
generated by cCQSs are highly complex and require infinite memory for optimal
prediction. That is, measuring a QSSP greatly obscures the underlying
quantum stochastic process. Fortunately, we have metrics to characterize these
processes and to develop a quantitative understanding of how measurement
affects the measured quantum processes. 

\subsection{Genericity of Complexity}
\label{sec:properties}

The tools are in place now to quantitatively analyze the measured QSSPs that
are represented by measured cCQSs. Here, we use the tools to draw broader
conclusions about what one should expect and how measurement choice changes the
randomness and complexity of measured quantum processes.

The main lesson from the preceding is that one expects explosive complexity
and this is reflected in the information-theoretic metrics of the measured
process.

{\Prop{A measured quantum process, with a measured cCQS presentation,
generically is highly complex in two specific ways: it has nonzero entropy
rate and statistical complexity dimension. That is, it requires uncountably
infinite states to optimally predict.
}}

{\ProProp{This follows as a corollary of Sec. \ref{sec:EmergentNonunifilar}'s
structural propositions---specifically Props.
\ref{prop:EmergentNonunifilarity} and \ref{prop:genericity} and Sec.
\ref{sec:Explosive}'s conjecture---though translated into the information
metrics of Sec. \ref{sec:Metrics}.
}}

As discussed above and extensively in Refs. \cite{Marz17a, Jurg20b, Jurg20c,
Jurg20e}, nonunifilar HMCs lead to causal state sets of uncountably infinite
cardinality and divergent statistical complexity. As the preceding
demonstrated, measured quantum processes have presentations that fall into
this class.

\section{Applications}
\label{sec:Apps}

Having laid out the progression from quantum sources to quantum state
processes and their presentations to measured processes and their metrics, we
are now ready to illustrate uses and benefits. The following does these via
three applications: measurement choice, alternate measurement protocols, and
optimal measurements.

\subsection{Measurement Variation and Choice}
\label{ssec:erVmmt}

Equations (\ref{eqn:param}) and (\ref{eqn:trans_probs}) directly show that
choice of measurement basis changes the observed process. This, in turn, means
that process' entropy rate and its MSP's statistical complexity dimension
also depend on measurement choice. Fortunately, the changes are well behaved.

{\Conj{Measured process complexity depends piecewise smoothly on both the underlying QSSP 
and choice of measurement. 
}
\label{conj2}
}

\begin{figure}[htbp]
\centering
\includegraphics{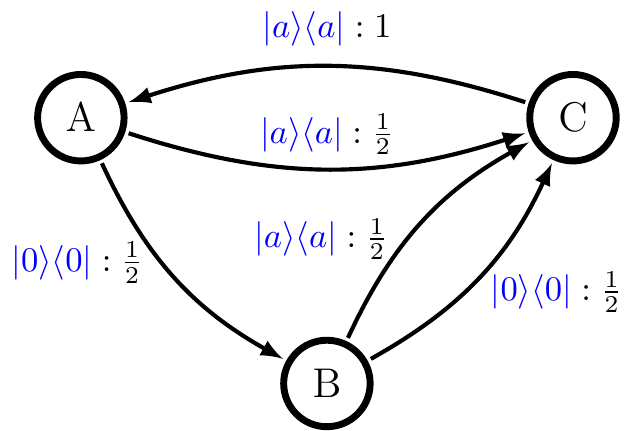} 
\caption{\qemach that generates a quantum process with qubits in quantum states
	$\ket{0}$ and $\ket{a} = \cos{\pi/5}\ket{0} + \sin{\pi/5}\ket{1}$.
	}
\label{fig:0a_rip}
\end{figure}

{\Rem{Given the extensive development up to this point, the following refrains
from presenting formal proofs. These will appear elsewhere. Nonetheless, it is
worthwhile to illustrate how the results can be used to outline a construction
that supports observed behavior and is backed by formal proofs in parallel
problem settings.

\begin{figure*}[htbp]
\hspace{0.5in}
\includegraphics[width=\textwidth]{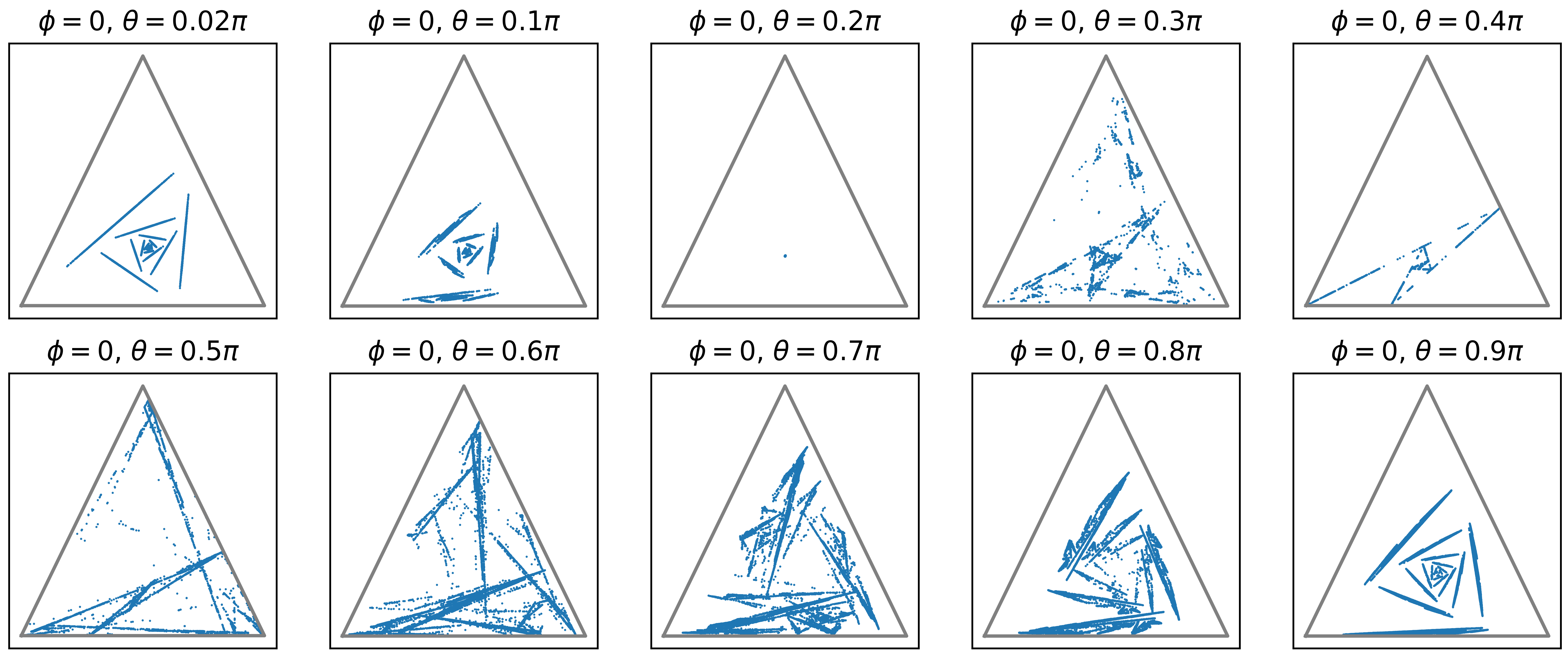}
\caption{Mixed-state presentation of the process resulting from measuring the
	quantum process generated by the \qemach in Fig. \ref{fig:0a_rip} in the
	measurement bases parametrized by $(\phi,\theta)$, as indicated in each
	subfigure. For each value of the parameters 30000 mixed states are plotted.
	}
\label{fig:rip_msps}
\end{figure*}

Note that: 
\begin{enumerate}
\item The measured process' entropy rate and statistical complexity dimension 
	depend smoothly on its MSP's invariant measure, as can be seen from Eqs.
	(\ref{eqn:hmuB_def}) and (\ref{eqn:dmu_def}). 
\item Equations (\ref{eqn:param}) and (\ref{eqn:trans_probs}) state that the
	parameters (transition probabilities) of the \measuredM HMC depend
	smoothly on the underlying QSSP and measurement operator parameters.
\end{enumerate}
Therefore, if the MSP's invariant measure depends smoothly on the parameters
of the \measuredM HMC, then the entropy rate and statistical complexity
dimension of the measured process depend smoothly on the underlying QSSP and
measurement parameters. 

Smoothness dependence of the MSP's invariant measure with respect to HMC
parameters is not only consistent with observation, which is illustrated
shortly, but has been established for many classes of \emph{iterated function
system} (IFS). For more detail, Ref. \cite{Jurg20b} outlines how any HMC can
be cast as an IFS---a stochastic dynamical system with a unique attractor
(equivalent to an HMC's MSP) that has an invariant measure. Both the attractor
and the invariant measure vary smoothly as a function of IFS parameters under
contractivity conditions \cite{Cent94a, Mend98a, Kloe20a}. These conditions
are generally satisfied by HMCs, thus indicating that both the MSP and
invariant measure of an HMC depend smoothly on the HMC parameters.
The caveat of piecewise smoothness as opposed to smoothness stems from the 
fact that the MSP can have abrupt jumps in cardinality for a finite set of parameters,
potentially causing finite discontinuities in the statistical complexity dimension, as will 
be illustrated in the examples of the following section.}}

Beyond smooth dependence, we ask more specifically, How do the mixed-state
invariant measure and the associated complexity measures change as a function
of the measurement angles $\theta$ and $\phi$? To answer these questions, we
explore two specific examples. For each we choose a quantum process generated
by a particular \qemach. We then obtain the measured cCQSs resulting from
measuring the quantum process with bases in which one of the angles is held
fixed and the other sweeps across its range of possible values.

The two example processes below were chosen since together they illustrate the
general properties of measurement dependence of QSSPs. The first example is
the three-hidden-state cCQS depicted in Fig. \ref{fig:0a_rip}, which is then
measured in many different qubit bases, holding $\phi=0$ and varying $\theta$.
The second example is the two-hidden-state quantum process generated by the
cCQS in Fig. \ref{fig:ex21}, which is then measured following the same
procedure.

\begin{figure}[htbp]
\centering
\includegraphics[width=\columnwidth]{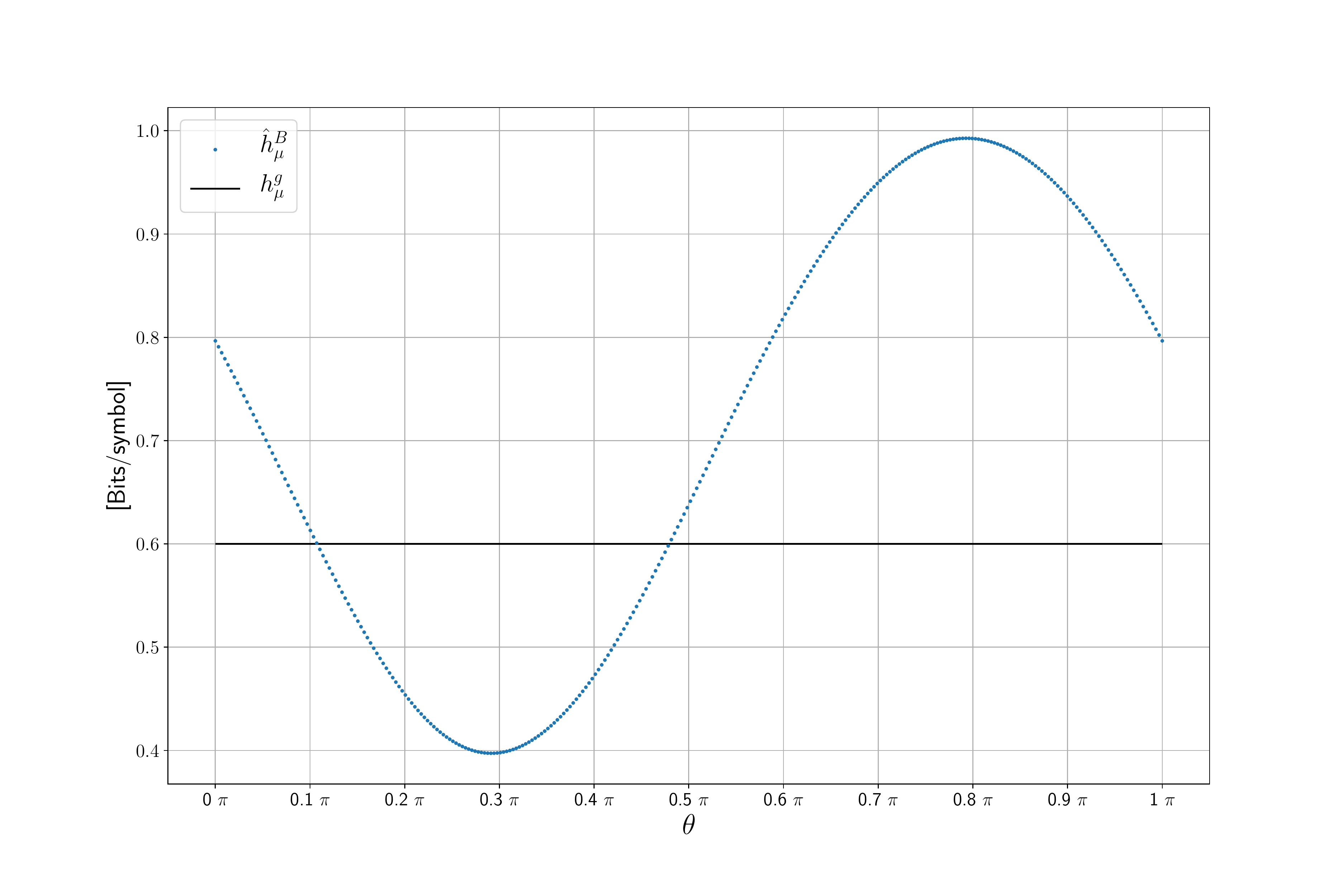} 
\caption{Entropy rate of the \measuredMs resulting from measuring the quantum
	process generated by the \qemach of Fig. \ref{fig:0a_rip} as a function of
	measurement angle $\theta$, as in Eq. (\ref{eqn:param}) at $300$ $\theta$
	values with the value $\phi =0$ fixed. Entropy rate $h_\mu^g$ (black line)
	of the \qemach that generates the measured process.
}
\label{fig:rip_hmus}
\end{figure}

\subsubsection{Random Insertion Process}
\label{sssec:rip}

First, we track changes in the invariant measure on the mixed-state simplex.
Figure \ref{fig:rip_msps} shows these for the measured process generated by
the \qemach of Fig. \ref{fig:0a_rip} in $10$ different measurement bases
$(\phi,\theta)$, as noted there. The structure of the invariant sets
$\MxSSet$ varies substantially with measurement basis. For most (but one,
discussed below) measurement bases the set has an uncountable infinity of
states, yet these states have distinct structures that vary smoothly with
choice of measurement basis. 

\begin{figure*}[htbp]
\hspace{0.5in}
\includegraphics[width=\textwidth]{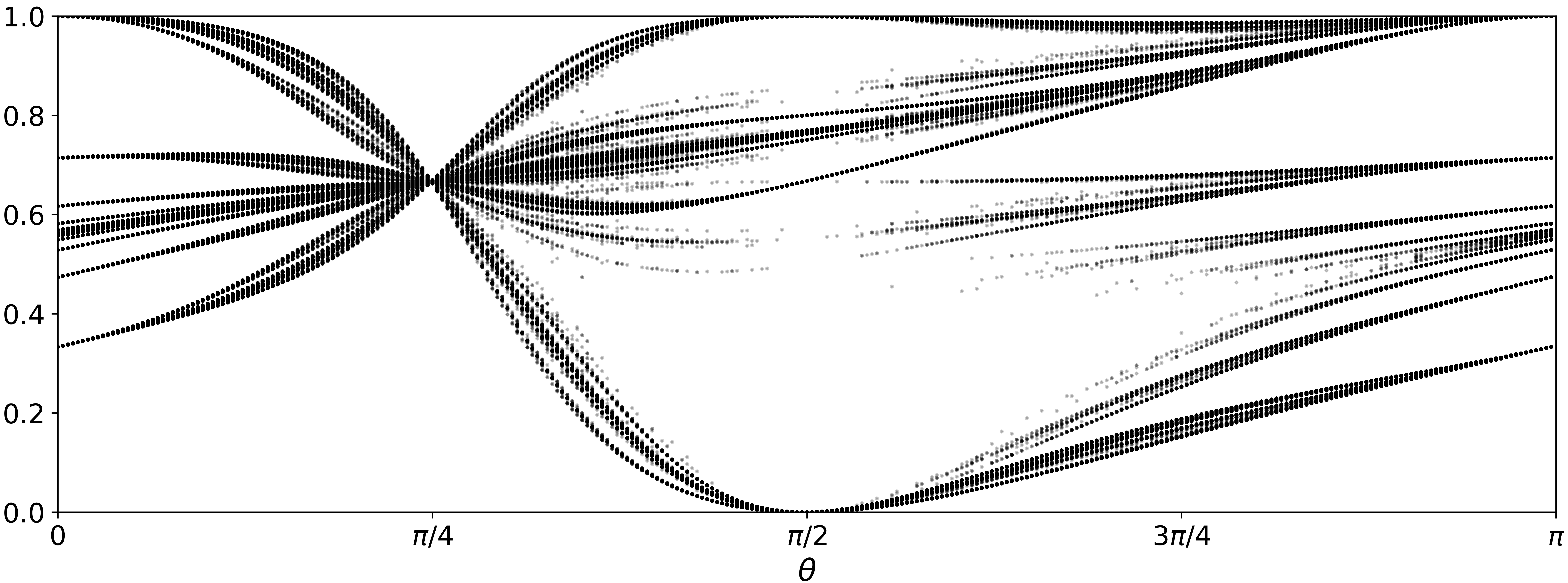}
\caption{Mixed-state presentations of the processes resulting from measuring the
	quantum process generated by the \qemach in Fig. \ref{fig:ex21} in the
	bases parametrized by $(\phi=0,\theta)$. Each vertical line represents the
	1D simplex $\MxSSet$ and the points in it are the mixed states corresponding
	to the measured cCQS at that particular value of $\theta$.
	}
\label{fig:gm_msps}
\end{figure*}

Second, we determine the entropy rate as a function of measurement basis in
Fig. \ref{fig:rip_hmus}. The process is measured in $300$ different bases,
holding the value of $\phi = 0$ and varying $\theta \in [0,\pi]$. By comparing
to the entropy rate $h_\mu^g$ of the \qemach that generates the underlying
QSSP, Fig. \ref{fig:rip_hmus} clearly demonstrates that measurement both
increases and decreases the randomness ($\hmu$).

While this example serves to graphically illustrate the high complexity of
predicting the classical stochastic processes measured from the QSSP, there
are limitations to estimating the MSP's statistical complexity dimension. For
reasons explained in detail in Ref. \cite{Jurg20c}, estimating the statistical
complexity dimension for measured cCQSs with MSPs in two and higher dimension
simplices is computationally intensive and there is as yet no efficient
algorithm. To illustrate the behavior of the statistical complexity dimension
in these stochastic processes, though, we turn to an example of a QSSP
generated by a two-state cCQS with an MSP in the $1$-simplex.

\subsubsection{Golden Mean Process}
\label{sssec:gm}

This example analyzes the QSSP generated by the cCQS in Fig. \ref{fig:ex21}.
It is measured in different qubit bases holding $\phi =0$ fixed and varying
$\theta$ uniformly from $0$ to $\pi$. Each measurement yields a measured cCQS
that is nonunifilar, the MSP is then computed. Figure \ref{fig:gm_msps}
displays its invariant measures. Each vertical unit interval corresponds to a
$1$-simplex that shows the MSP at that particular value of $\theta$. From the
figure we observe that the majority of the MSPs have a complex fractal-like
structure. However, what the figure makes evident is that this structure
varies smoothly with respect to the measurement parameter $\theta$, consistent
with Sec. \ref{ssec:erVmmt}'s Conjecture. 

\begin{figure}[htbp]
\centering
\includegraphics[width=\columnwidth]{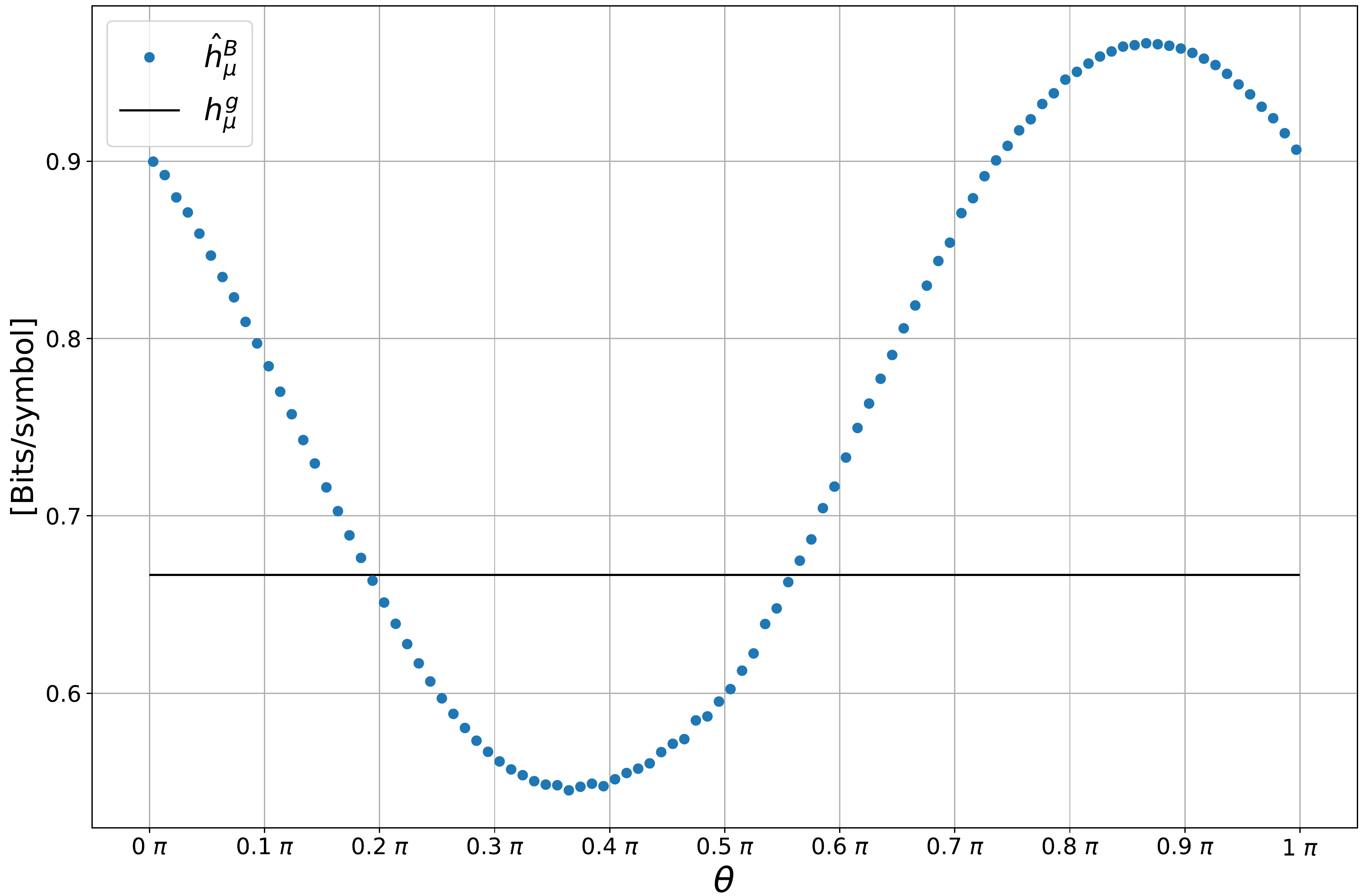} 
\caption{Entropy rate of the \measuredMs resulting from measuring the quantum
	process generated by the \qemach of Fig. \ref{fig:ex21} as a function of
	measurement angle $\theta$, as in Eq. (\ref{eqn:param}) with the value
	$\phi =0$ fixed, at $100~\theta$ values. Entropy rate $h_\mu^g$ (black
	line) of the \qemach that generates the underlying QSSP.}
\label{fig:gm_hmus}
\end{figure}

\begin{figure}[htbp]
\centering
\includegraphics[width=\columnwidth]{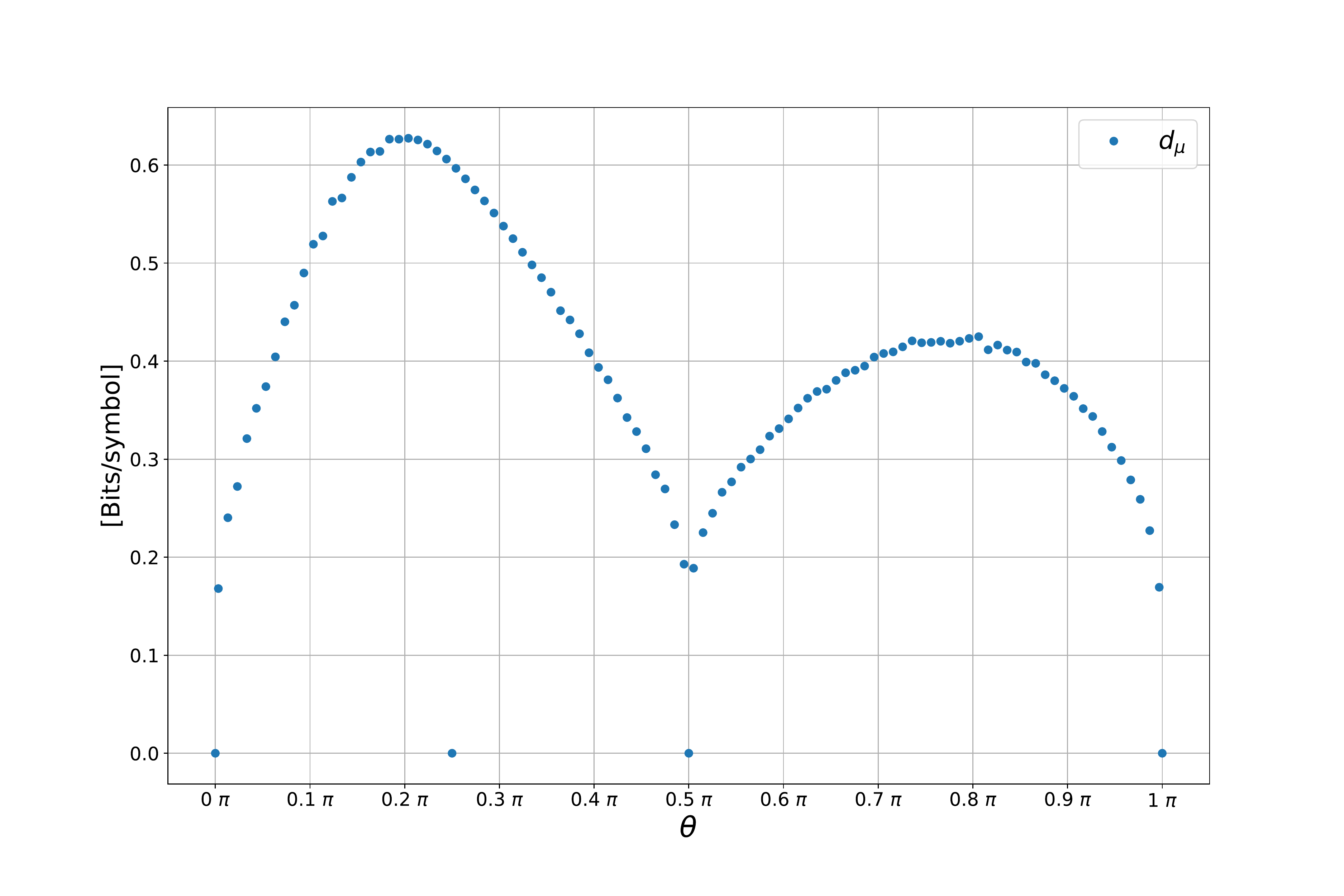} 
\caption{Statistical complexity dimension $d_\mu$ of the \measuredMs resulting
	from measuring the quantum process generated by the \qemach of Fig.
	\ref{fig:ex21} as a function of measurement angle $\theta$, as in Eq.
	(\ref{eqn:param}) with the value $\phi =0$ fixed, at $100~\theta$ values.
	}
\label{fig:gm_dmus}
\end{figure}

Figures \ref{fig:gm_hmus} and \ref{fig:gm_dmus} track how both entropy rate
$\hmu$ and statistical complexity dimension $d_\mu$ vary with respect to
measurement basis.

As with the previous example, we see that $\hmu$ of the measured
process both increases and decreases with respect to $h_\mu^g$---that of the
original QSSP---depending on measurement basis.

The statistical complexity diverges for most, in contrast with the finite
statistical complexity of the underlying QSSP. That said, the statistical
complexity dimension $d_\mu$ smoothly varies. To a certain extent this
reflects what is see from the MSPs in Fig. \ref{fig:gm_msps}. Figure
\ref{fig:gm_dmus} also reveals four values of $\theta$ for which $d_\mu = 0$.
These are are $\theta \in \{0, \pi/4, \pi/2, \pi \}$. When $\theta$ takes the
values $0$ or $\pi$ the measurement is in the observation basis and one of the
measurement operators aligns with the quantum state $\ket{0}$. This simplifies
the process and thus the \measuredM has a countably infinite number of mixed
states. This also happens at the value of $\theta = \pi/2$, in which one of
the measurement operators aligns with the quantum state $\ket{+}$ that is
output by the cCQS. Notice that in these three cases, $d_\mu =0$ is reached
smoothly. 

The exception to smoothness is the discontinuous jump to $d_\mu = 0$ when the
process is measured at $\theta = \pi/4$. This special case is discussed more,
shortly. In general terms, though, for that particular basis the measurement
does not distinguish between the two distinct emitted qubit states $\ket{0}$
and $\ket{+}$. And so, all of the structural information about the underlying
quantum process is lost, except for the probability of obtaining one
measurement outcome or the other. The process becomes memoryless and
so has a single-state presentation.

\subsubsection{General Features}
\label{sssec:feats}

With the previous two examples in hand, and after an exhaustive exploration of
example processes in this fashion, we review several common characteristics.
Of particular interest are the smooth behaviors of $\hmu$ and $d_\mu$ with
well defined maxima and minima. It is also apparent that the MSP invariant
sets exhibit marked structural variations. However, in agreement with Sec.
\ref{ssec:erVmmt}'s Conjecture, they appear to vary smoothly with respect
to measurement change. 

A feature that immediately warrants attention in Fig. \ref{fig:rip_msps} is
the drop in structural complexity of the MSP at $\theta = \pi/5$. With that
particular measurement basis, the statistical complexity dimension vanishes,
indicating that the measured cCQS is finite. On closer inspection, the HMC
corresponding to the measured cCQS is not only finite, but has a single causal
state. This indicates that the measured process consists of independent
identically distributed (i.i.d.) random variables. At each time step, the
observed symbols are $0$ with probability $p_0 = \cos^2{\pi/5}$ and $1$ with
probability $p_1 = \sin^2{\pi/5}$. This seemingly special case is not a fluke.

{\Prop[Memoryless measurements]{For any cCQS with quantum alphabet
$\mathcal{A}_Q$ consisting of two distinct quantum states $\rho_a$ and
$\rho_b$, there exists a set of measurement bases for which the resulting
measured process is memoryless and $\Cmu = 0$.
}}

{\ProProp{We establish this by construction. Without loss of generality and
for ease of notation we align both quantum states with the $zx$-plane, such
that one of the quantum states $\rho_a$ is at the top of the Bloch sphere. We
further denote the angle between the two states by $\alpha$. We then write
$\rho_a = \ket{0}\bra{0}$, and $\rho_b = \ket{b}\bra{b}$ such that $\ket{b} =
\cos{\alpha/2}\ket{0} + \sin{\alpha/2}\ket{1}$.

Consider the projective measurement bases for which one measurement operator
projects onto a state $\ket{\psi_0}$, such that $|\braket{\psi_0|0}| =
|\braket{\psi_0|b}|$. That is, $\ket{\psi_0}$ lies in the Bloch sphere
circumference that bisects the angle between $\ket{0}$ and $\ket{b}$.  Then,
the set of measurements that project onto $\ket{\psi_0}$ and
$\ket{\psi_0^\perp} \equiv \ket{\psi_1}$ are such that the probability
distributions over measurement outcomes are the same whether the measured
qubit state was $\rho_a$ or $\rho_b$. That is, for this particular set of
measurements, we have $\Pr(i|\rho_j) = p_i$ for $i \in \{0,1\}$ and for all
$\rho_j$ with $p_i$ a constant and $p_0+p_1=1$.

Together with Eq. (\ref{eqn:trans_probs}), this observation says that the
\measuredM transition matrices are, for $i \in \{0,1\}$: 
\begin{align}
T^i & =  \mathbb{T}^{\rho_a} Pr(i|\rho_a) + \mathbb{T}^{\rho_b} Pr(i|\rho_b)
	\nonumber \\
    & = p_i (\mathbb{T}^{\rho_a} + \mathbb{T}^{\rho_b}) 
	\nonumber \\
	& = p_i \mathbb{T}
  ~.
\label{eqn:labeled_Ts}
\end{align}
That is, both labeled transition matrices are proportional to each other and
to $\mathbb{T}$. Note also that $T = T^0 + T^1 = \mathbb{T}$, so for
simplicity we refer to the internal Markov chain transition matrix as $T$. 

Both labeled transition matrices being proportional to T implies that the MSP
yields a biased coin process with biases $p_0$ and $p_1$, respectively. There
is a single recurrent mixed state, namely $\pi$. This follows by definition,
since $\pi$ is an eigenvector of $T$. And so, evolving the mixed state $\eta_t
= \pi$ gives: 
\begin{align*}
\eta_{t+1} & = \frac{\pi \cdot T^i}{p_i} \\
  & = \pi \cdot T \\
  & = \pi
  ~.
\end{align*} 
}}

Physically, memoryless measurements project states onto a basis whose
components are symmetric with respect to the pure states in $\mathcal{A}_Q$.
Consequently, the measurement cannot distinguish between the pure states and
so the act of measurement effectively leads to a complete loss of information
about the cCQS's internal structure.

The fact that these memoryless measurements maximize the loss of information
about the cCQS's internal structure, naturally leads to the question of
whether there exists a set of measurements that maximally preserves
information about the cCQS's internal structure. These would be measurements
that optimally distinguish between the quantum states. In the case of POVMs
these measurements are well studied for the case of distinguishing between $2$
or $3$ states. And, as explored in Sec. \ref{sec:povms} they in fact yield
special measured processes. 

Shortly, we return to explore the issue of optimal and extremizing measurement
bases.

Another point to make is that in all of the examples above, only the $\theta$
parameter of the measurement bases was varied, and the phase $\phi$ was fixed
to zero. This choice was for simplicity and visualization purposes only. The
variation of $\phi$ does not change the analysis or the conclusions in any way.
The only notable point is that when the measurement parameters both define a
measurement basis that is not close to being aligned with any of the output
quantum states and is poor at distinguishing between them, then that will in
general result in processes that have $h_\mu$ larger than the underlying
\qemach, and as is generally the case, divergent memory.

Appendix \ref{app:mad} graphically demonstrates this with two animations that
sweep the angles $\theta$ and $\phi$ while monitoring entropy rate and mixed
states. One animation shows how the mixed state presentation and $h_\mu$ vary
a function of parameter $\theta$. The other animation shows $h_\mu(\theta)$
plots as in Fig. \ref{fig:rip_hmus} while sweeping $\phi$ from $0$ to $2\pi$.

In general, as seen from Fig. \ref{fig:rip_hmus}, different choices of
measurement increase or decrease the randomness of the measured quantum
process. Furthermore, even if the general case is that quantum measurement
dramatically increases the structural complexity of the measured stochastic
process with respect to the underlying quantum process, the existence of
memoryless measurements shows that particular choices of measurement, in
fact, can mask the quantum process' structural complexity. 

\subsection{Alternate Measurement Protocols}

While simplicity dictated that the preceding concentrate on protocols in which
the same projective measurement is applied at every time step, there are many
alternative protocols to explore. As an example, the following considers
processes that result from applying more general measurements to each emitted
qubit.

\subsubsection{Positive Operator-Valued Measurements}
\label{sec:povms}

The development to this point investigated the consequences for the observed
classical stochastic process of employing only projective measurements.
However, the natural generalization is to more flexible positive
operator-valued measurements.

{\Def[Positive Operator-Valued Measurement on a QSSP]{A
\emph{positive operator-valued measurement} (POVM) $\mathcal{I}$, consists of a finite set of 
positive semi-definite operators $\{E_x\}$, on the Hilbert space $\mathcal{H}^d$ 
of dimension $d$. The operators satisfy the condition $\sum_x E_x = \mathbb{I}_d$.
When measurement $\mathcal{I}$ acts on a quantum system in state $\rho_t$,
emitted
by a QSSP, the outcome is $x_t$, corresponding to operator $E_x$, with
probability:
\begin{align*}
\Prob(x_t|\rho_t) = tr(\rho E_{x_t})
  ~.
\end{align*}
Applying a POVM to every quantum state emitted by the QSSP yields a classical
stochastic process over the values of $x \in \mathcal{A}_M$---the alphabet of
the measured process.
}}

When the measurement $\mathcal{I}$ consists of a POVM, the number of operators
$\{E_j\}$ and possible outcomes can be any positive integer. This increase in
possible measurement outcomes results in a larger alphabet for the classical
measured quantum process. At first glance, this suggests finding a wider range
of unifilar measured HMCs, but it is not. This is a direct result of
indistinguishability.

Generally, when performing a POVM on any two qubit states (even
distinguishable ones), at least one of the measurement outcomes has a nonzero
probability of being observed on both of the quantum states. This is due to
the fact that POVMs generally have nonorthogonal measurement operators. 

When measuring with a POVM, consider a cCQS hidden state with two outgoing
transitions on distinct quantum states. Applying the POVM on those transitions
means that at least one of the symbols in the classical alphabet
$\mathcal{A}_M$ is present in two outgoing transitions for the same
hidden state in the \measuredM. This makes the dynamic of the \measuredM
nonunifilar. Thus, in general when using POVMs, measured processes are also
highly complex, akin to those obtained when applying projective measurements.

That said, the measured processes produced using POVMs reveal a new collection
of notable special cases, which remain to be broadly explored. To illustrate
just one, the following develops a simple but illuminating example using the
unambiguous state discrimination POVM. The resulting flexibility then leads,
in the following section, into the challenge of optimizing measurement
protocols to achieve various ends. 

\subsubsection{Unambiguous State Discrimination}

Recall that when measuring qubits either in state $\ket{\psi}$ or $\ket{\phi}$, the POVM yielding the highest probability of unambiguously distinguishing between them is given by \cite{Ivan87a, Diek88a, Pere88a}:
\begin{subequations}
\begin{align}
E_{\psi} & =
  \frac{1}{1+|\braket{\phi|\psi}|}\ket{\phi^{\perp}}\bra{\phi^{\perp}} \\
E_{\phi} & =
  \frac{1}{1+|\braket{\phi|\psi}|}\ket{\psi^{\perp}}\bra{\psi^{\perp}} \\
E_? & = \mathbb{I} - E_{\psi} - E_{\phi}
  ~.
\end{align}
\label{eqn:povm}
\end{subequations}

\begin{figure}[htbp]
\centering
\includegraphics{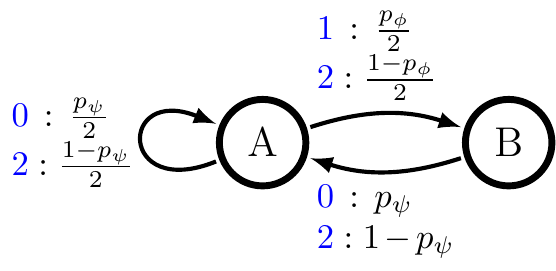}
\caption{HMC presentation of the process resulting from measuring the QSSP
	depicted in Fig. \ref{fig:qem0} with the POVM in Eq. (\ref{eqn:povm}).
	}
\label{fig:povm}
\end{figure}

Applying this measurement scheme to the quantum process generated by the
\qemach in Fig. \ref{fig:qem0} produces the classical process emitted by the
HMC of Fig. \ref{fig:povm}. There $\{E_\psi, E_\phi, E_?\}$ are relabeled
$\{E_0, E_1, E_2\}$, $p_\psi = \Tr(E_0\ket{\psi}\bra{\psi})$, and $p_\phi =
\Tr(E_1\ket{\phi}\bra{\phi})$.  

Note that symbol $2$, corresponding to an inconclusive measurement, is present
in all transitions. Yet observing $0$ or $1$ is synchronizing since they
each determine the next HMC state. This property is preserved from the
\qemach that is being measured and need not occur generally. That said, if the
\qemach under study outputs only two distinct quantum states, then measuring it
with the unambiguous state discrimination POVM in Eq. (\ref{eqn:povm}) results
in an HMC presentation that preserves the internal topology. However, each HMC
transition is corrupted with a nonzero probability of observing symbol $2$,
rendering an inconclusive measurement. 

\begin{figure}[htbp]
\centering
\includegraphics[scale=0.9]{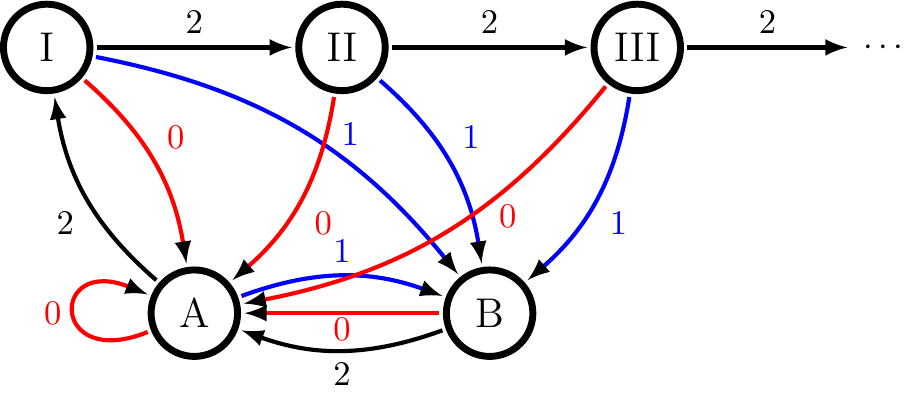} \caption{State transition diagram of the MSP constructed from the HMC in Fig.
	\ref{fig:povm}. Transition probabilities omitted for clarity. Observing
	$1$s (blue transitions) leads to state $B$; observing $0$s (red
	transitions) to state $A$. Both cases are synchronizing.
	}
\label{fig:povm_msp}
\end{figure}

For this example, constructing the MSP for the process generated the \qemach
in Fig. \ref{fig:povm} produces the presentation depicted in Fig.
\ref{fig:povm_msp}, where transition probabilities are not shown to reduce clutter.

There is a subset of MSP states with topology similar to the original \qemach,
but augmented with the mixed states that capture the observation of $2$s. 
This generally holds when the generator is
a unifilar \qemach that emits two distinct nonorthogonal quantum states if the
process is measured via the POVM in Eqs. (\ref{eqn:povm}). In the \measuredM,
there will be a subset of MSP states that mimic the \qemach's internal
dynamics, but the latter is augmented by inconclusive measurement
outcomes. And, there are chains of mixed states that drive the process away
from the original dynamic whenever a sequence of ``inconclusive results'' or a
nonsynchronizing symbol is observed. 

To explore this process further set:
\begin{align*}
\ket{\phi} & = \ket{0} \\
  \ket{\psi} & = \cos{(\alpha/2)}\ket{0} + \sin{(\alpha/2)}\ket{1}
  ~,
\end{align*}
with $\alpha \in (0, \pi)$.
Then, $p_\psi = p_\phi = 1 - \cos{(\alpha/2)}$. When constructing the MSP of Fig.
\ref{fig:povm_msp}, all transitions that emit a $2$ (black) have an associated
probability of $\cos{(\alpha/2)}$, while the probabilities of the blue and red
transitions depend on the specific transition. As the number of $2$s observed 
approaches infinity, the mixed states visited approach the stationary state 
distribution $\pi = (2/3, 1/3)$ of the nonunifilar \measuredM.

\begin{figure}[htbp]
\includegraphics[clip,width=\columnwidth]{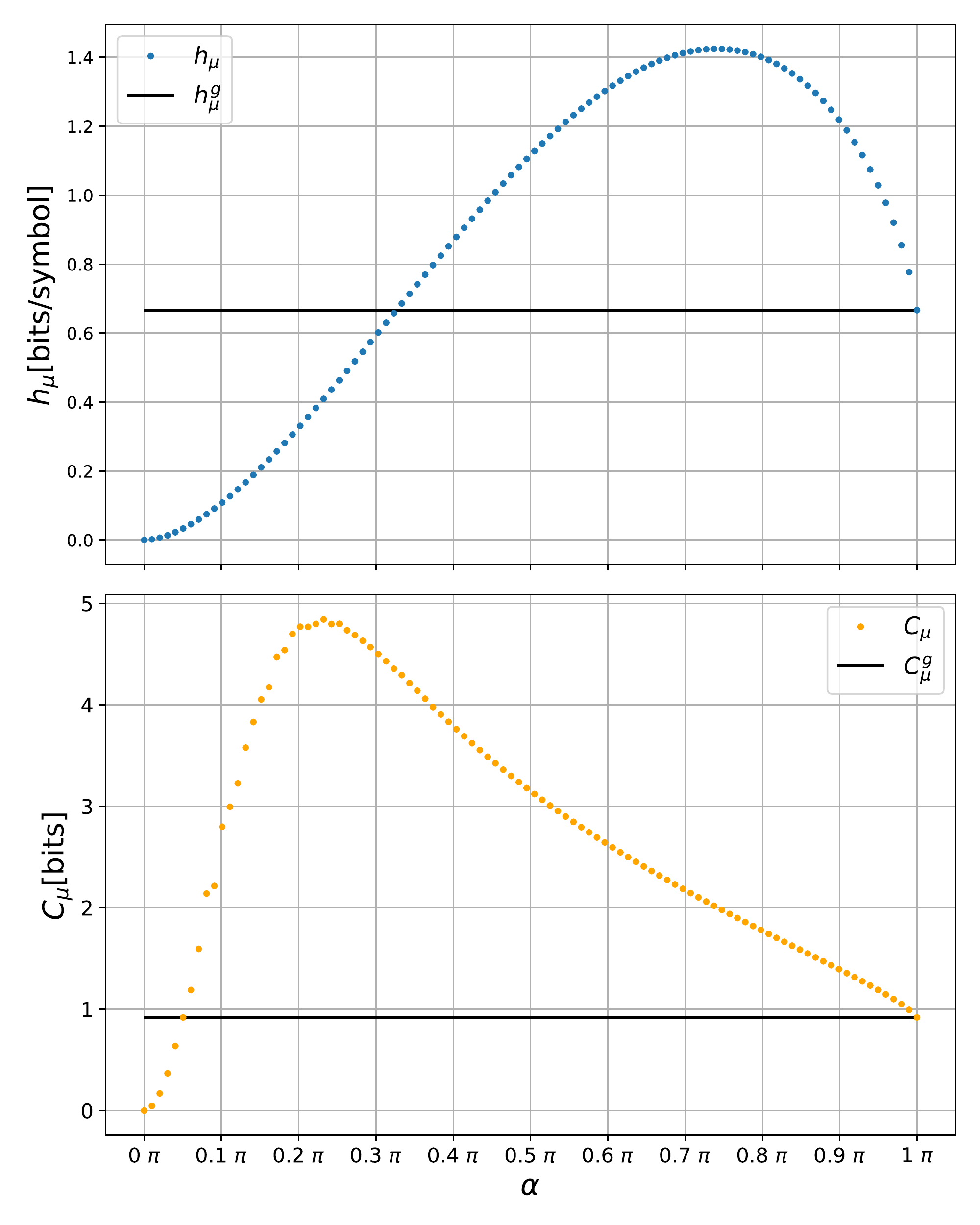}\caption{(Top) Randomness $\hmu$ and (bottom) statistical complexity $\Cmu$ as
	a function of the angle $\alpha$ between the two states emitted by the
	\qemach in Fig. \ref{fig:qem0}, with $\ket{\phi} = \ket{0}$ and $\ket{\psi}
	= \cos({\alpha/2})\ket{0} + \sin({\alpha/2})\ket{1}$. The horizontal black
	lines show the values of $\hmu^g$ and $\Cmu^g$ of the \qemach that
	generates the original quantum state process.
	}
\label{fig:msp_povm_metrics}
\end{figure}

\begin{figure}[htbp]
\includegraphics[clip,width=\columnwidth]{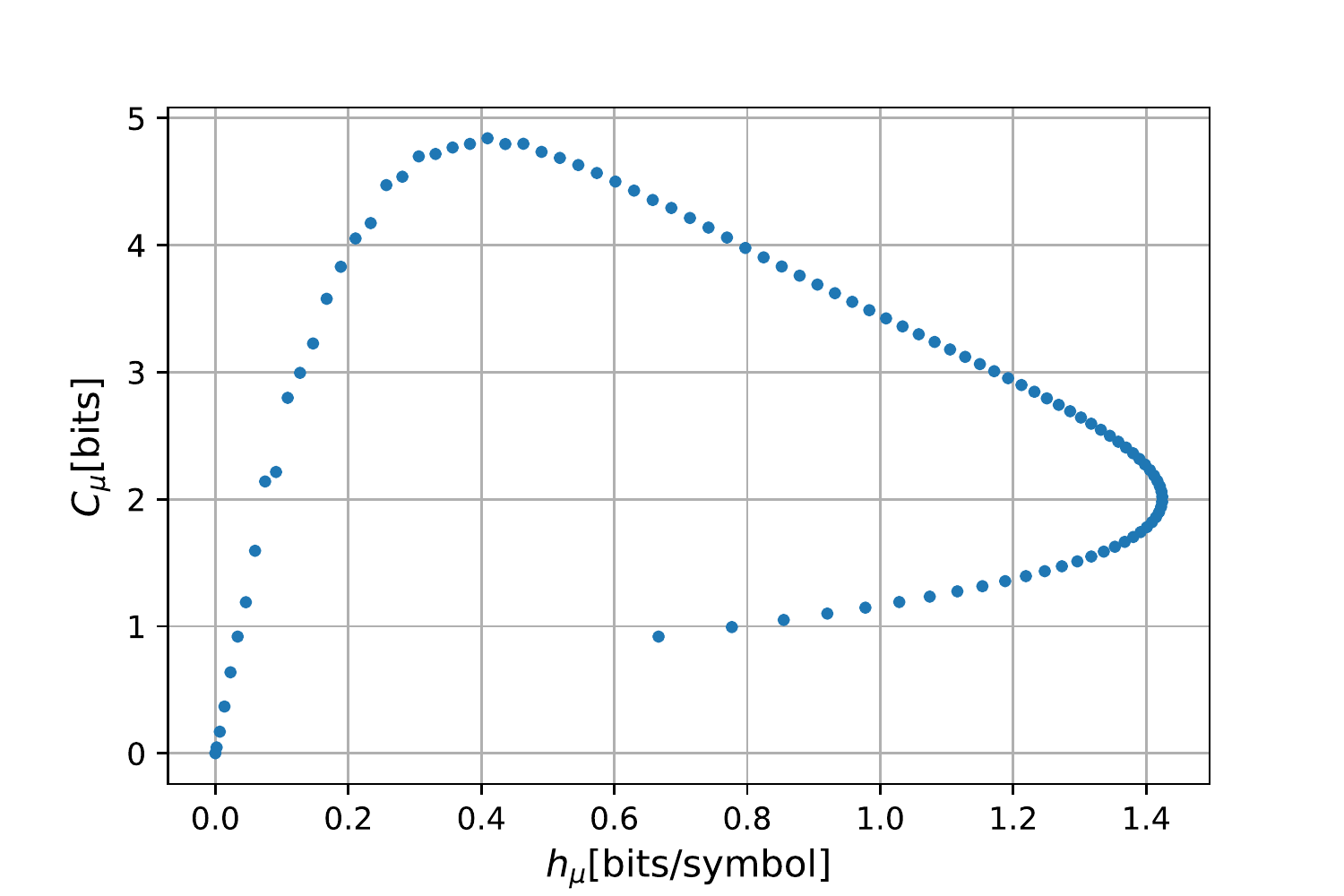}\caption{Complexity-entropy diagram capturing the purely informational
	character of the measured quantum-state process: Parametric plot of
	$\hmu(\alpha)$ and $\Cmu(\alpha)$ over $\alpha \in [0,\pi]$ illustrating
	how the intrinsic informational properties depend on each other without
	reference to model parameter $\alpha$. Cf. complexity-entropy plots in Ref.
	\cite{Feld08a}.
	}
\label{fig:comp_ent_povm}
\end{figure}

This MSP, while requiring a countable infinity of states, is so well behaved
that it allows for direct calculation of the mixed states and the numerical
computation of both its entropy rate and statistical complexity by
approximating the MSP with a finite but sufficiently large set of mixed states.
Figure \ref{fig:msp_povm_metrics} plots both the entropy rate and statistical
complexity when the processes are approximated by a MSP with $500$ hidden
states. 

Figure \ref{fig:msp_povm_metrics} reveals edge cases that match expectations.
At one extreme, when $\alpha = 0$, the process reduces to a sequence of qubits
in state $\ket{0}$. Thus, both process randomness and structure vanish. At the
other extreme, when $\alpha = \pi$, the alphabet emitted by the \qemach is
orthogonal $\{ \ket{0}, \ket{1} \}$. Hence, the unambiguous state
discrimination POVM reduces to the projective measurement aligned with the
observation basis. This means that the measured process is true to the original
quantum source and both its entropy rate and statistical complexity coincide
with $h_\mu^g$ and $C_\mu^g$, respectively. The plots also highlight that both
entropy rate and statistical complexity coincide with the generator values with
different nonorthogonal alphabets and that their maximum values are attained
for different alphabets as well. While both randomness and structure of the
measured process depend on the \qemach's quantum alphabet, they both have
distinct meanings and, thus, the dependencies are not equivalent.

When varying the quantum alphabets and exploring $\alpha$'s whole range, it
becomes apparent that the values of both entropy rate and statistical
complexity can be lower or higher than those of the original \qemach. This
means that for a given \qemach and a given measurement both the randomness and
structure of the measured process with respect to the QSSP can be reduced or
increased. This makes plain the possibly ambiguous effects of measurement and
what the latter can add to or remove from the underlying quantum process.

Figure \ref{fig:comp_ent_povm}'s \emph{complexity-entropy diagram}
\cite{Feld08a} offers a more concise display of the QSSP's achievable
information generation and storage---its \emph{intrinsic computation}---when
measured in this particular POVM across the range of alphabets. Notice that for
small $\hmu$ the system's structure or memory requirements $\Cmu$ are low. Then
memory increases with increased randomness until a peak is reached at about
$\hmu = 0.4$ bits/symbol. Above this, increased randomness requires fewer
memory resources and a given randomness can be achieved at more than one memory
value $\Cmu$.

Overall, this example illustrates a situation in which a particular choice of
measurement protocol leads to a very tractable measured process. While the
statistical complexity diverges for most measured processes, this example
shows that tailoring measurement schemes still leads to complex, but more
tractable dynamics. That is, the observed dynamics can be leveraged to better
understand the dynamic that produces the underlying QSSP.

\section{Optimal Measurements}
\label{sec:Optimal}

The previous sections established the two main characters of measured
quantum processes---their unpredictability and temporal correlation. And, they
demonstrated how measurement can increase or decrease observed randomness and
structure. These metrics naturally broach the challenge of defining and
demonstrating the existence of informationally-optimal measurements. The
possibility of these optimizations is greatly facilitated by the piecewise
smooth dependence of the informational metrics on the QSSP and on measurement
operators.

Eschewing details, the following lays out several avenues for future
exploration, illustrating various kinds of optimality using the tools now in
hand. We consider, in turn, measurements that lead to minimal structural
complexity and to various forms of maximal informativeness. The following cases
only address projective measurements, though extension to POVMs is in some
cases straightforward and of interest in general. 

Developing algorithms and calculational methods to find and implement these
optimal measurements is left to the future.

Let's briefly recall relevant notation. A given projective measurement
protocol is denoted $\mathcal{E}$. Given a QSSP $\QSSProcess$ and a
measurement protocol $\mathcal{E}$, the corresponding measured process is
$\MProcess = \mathcal{E}(\QSSProcess)$. To simplify the notation the following
introduces $\mathcal{E}(M)$---the MSP of the \measuredM of $\MProcess$ for a
given \qemach $M$. 

The following treats the alternative informational metrics as operators
themselves. So that for HMC $M$, $\hmu(M)$ is that HMC's entropy rate,
$\Cmu(M)$ is its statistical complexity, and $d_\mu(M)$ its statistical
complexity dimension.

\subsection{Minimal Structural Complexity}

There are settings where it is useful to identify and use measurements that
lead to the least complex, smallest-memory observed process. Such measurement
schemes are specified as follows.

{\Def[Minimal Structural Complexity Measurement]{Given a \qemach $M$, the
projective measurement $\mathcal{E}_{\underline {\Cmu}}$ that leads to the measured 
process with the minimal structural complexity is, when $\mathcal{E}(M)$ is finite state:
\begin{align*}
\mathcal{E}_{\underline{\Cmu}} = \argmin_{\{\mathcal{E}\}} \Cmu( \mathcal{E}(M) )
\end{align*}
and when $\mathcal{E}(M)$ uncountably infinite state:
\begin{align*}
\mathcal{E}_{\underline{\Cmu}} = \argmin_{\{\mathcal{E}\}} d_\mu ( \mathcal{E}(M) )
  ~.
\end{align*}
}}

While this kind of measurement is the least informative about the underlying
quantum dynamics, it has also proven in multiple examples to be the measurement
that yields a classical process requiring the least memory resources to
simulate and predict. This remains to be proven but is consistent with the fact
that the measurement effectively is the most efficient at discarding
information about the structure of the underlying process, which need not be
stored to represent the resulting measured process. 

\subsection{Maximally Informative Measurements}

Perhaps most naturally, one can employ a measurement scheme that maximizes the
amount of information per symbol in the measured process.
Such measurement schemes are specified as follows.

{\Def[Maximally Informative Measurement]{Given a \qemach $M$, the
projective measurement $\mathcal{E}_{\overbar{\hmu}}$ that leads to the measured
process with maximally informative measurement outcomes is:
\begin{align*}
\mathcal{E}_{\overbar{\hmu}} = \argmax_{\{\mathcal{E}\}} \hmu( \mathcal{E}(M) )
  ~.
\end{align*}
}}

Recalling basic dynamical systems, this is a natural choice of optimal
measurement in that it mimics the essence of what a generating partition is,
as defined by Kolmogorov and proven by Sinai \cite{Sina59}. In this case, each
observation of a measurement outcome $X_t$ results in the maximum possible
amount of new information. 

\subsection{Maximally Mutually-Informative Measurements}

One is often interested in monitoring how measurement outcomes reveal (or not)
the internal generating mechanism. This suggests the following measurement.

{\Def[Maximally Mutually-Informative Measurement]{Given a QSSP $\QSSProcess$,
the measurement protocol $\mathcal{E}_{\overbar{R:X}}$ is \emph{the maximally
mutually informative measurement} when: 
\begin{align*}
\mathcal{E}_{\overbar{R:X}} = \argmax_{\{\mathcal{E}\}} I(\QSSProcess :  \MProcess)
  ~,
\end{align*}
where $I(\QSSProcess :  \MProcess)$ is the mutual information \cite{Cove91a}
between the QSSP and the measured process. 
}}

This measurement maximizes the information shared between the quantum-state
stochastic process and the measured quantum process. That is, observation of
the classical process maximally reduces uncertainty of the quantum process.  In
contrast to the \emph{maximally informative measurement}, the \emph{maximally
mutually-informative measurement} does not yield the maximal amount of
information learned per observation of the measured process. Rather, it
provides the maximal amount of information learned about the QSSP given
observation of the measured process.

\subsection{Dynamically Informative Measurements}

Finally, one may be interested in finding the measurement that yields a
stochastic process most similar to the underlying QSSP. This requires adhering
to a particular definition of distance or similarity between stochastic
processes. One option when working with a particular \qemach and its
corresponding \measuredM is to use a measure of distance between HMCs. The
measurement that minimizes the distance between two HMCs is the most
informative about the internal structure of the QSSP generated by the \qemach.
Said simply, it is the measurement yielding a classical stochastic process that
is most informative about the QSSP's dynamical structure. 

Selecting an appropriate measure of distance between HMCs is not a
straightforward problem. Many have been proposed \cite{Falk95a, Sahr10a,
Zeng10a}, each with their own nuances. Determining which distance measure
better suits the problem at hand is left for future work.

This and the above notions of optimality are distinct and so are of interest in
different operational settings. It is important to emphasize that, what
differentiates these optimality criteria from other notions of measurement
optimality is that they depend on the QSSP's time correlations and not only on
the particular quantum state of a single quantum system or its evolution.

\section{Conclusion}
\label{sec:Conclusion}

To investigate temporal complexity---unpredictability and structure---in
quantum dynamics we developed an intentionally simplified setting---one that
excluded sequential qubit entanglement. This allowed deploying classical
multivariate information theory as a quantitative analysis tool. And, this led
directly to isolating the problem of how measurement affects the appearance of
quantum processes---processes to which one must apply a quantum measurement to
observe. The simple lesson is that measurement can both increase or decrease
randomness and structure. In point of fact, and somewhat unanticipated,
observing a quantum process through projective measurements results in an
observed classical process of explosively high structural complexity. The
detailed analysis enriched this by identifying the mechanisms through which
this complexity arises.

In general, quantum-state stochastic processes observed through projective
measurements result in observed classical processes that require storing an
infinite number of predictive features to allow for optimal prediction. The
sets of predictive features for most processes are rich in structure and,
making use of that, we implemented newly developed tools to quantify their
structural complexity and the intrinsic randomness of the measured process.  
In addition, the development shed light on the influence that the chosen
measurement basis has on the complexity of the observed process.

Irreducible nonunifilarity was identified as the driving mechanism of these
features. The low dimensionality of the quantum state Hilbert space is the
physical cause. Nonalignment between the measurement basis and emitted quantum
states of the cCQS is the root physical mechanism that leads to induced
indistinguishability of quantum states.

Even allowing for the framework's simplifications, the typical complexity of
the measured process complicates not only its study, but also makes the task of
learning about the underlying quantum process a difficult one. We made progress
in understanding why that is and how to characterize the measured processes.
That progress came from adapting new methods from ergodic theory and random
dynamical systems to this setting. The result is a powerful toolset for
quantitatively analyzing measured quantum processes.

We showed that the underlying cCQSs have distinct signatures of structure and
randomness as a function of the measurement parameters and that dependence is
systematic and smooth. One remaining task is to characterize the possible
underlying quantum sources that generate a given measured cCQS or, at the very
least, their statistical properties.

In this way, the results lay a path to fully characterizing quantum state
stochastic processes. However, many steps remain unexplored. We conjecture that
success in these will have broad impacts. One of those steps is to find the
spectral decompositions of the processes by use of meromorphic functional
calculus \cite{Riec18a}. While these tools are not engaged here, they will be
necessary when studying CQSs generated by purely quantum controllers. Another
essential step is to model the internal controller in such a way that it
generates entangled QSSPs. There are helpful starting points for this in both
finite-memory classical controllers \cite{Cros08a} and quantum protocols
\cite{Scho05a, Scho07a} for sequential generation of matrix product states.
The latter are of particular interest in the study of many-body entangled
states. 

We only briefly explored a measurement protocol that used single qubit POVM
measurements. This showed that exploring different measurement protocols has
the potential to bring novel results and to move closer to more physically
realistic settings. For example, Ref. \cite{Gier21a} looks at measurements that
allow synchronization to the underlying QSSP in a setting similar to that
explored here. As in the study of classical dynamical systems, though,
understanding the informational and statistical effects of choosing a
particular measurement instrument or protocol can aid in optimizing particular
tasks. The study of optimal quantum measurements for QSSPs remains as a
challenging open problem.

A major challenge is to extend the current setting to quantally-controlled
qubit sources (qCQS), as just noted. And, then, from there to develop a
quantum communication channel setting in which qubits are input, quantally
processed, and then output. Advances in this will more directly impact
information processing and computing performed by quantum dynamical systems.
Beyond that, qubit source timing issues should be addressed, moving away from
the admittedly simple use of discrete time here to continuous-time processes.
Fortunately, the cCQS model can easily be extended to discrete-event
continuous-time hidden semi-Markov models using the methods of Ref.
\cite{Marz17b}. This will immediately give metrics of quantum randomness and
structure, paralleling the development here. 

Extending the present results along these lines will naturally complement
existing quantum descriptions of classical stochastic processes \cite{Gu12a,
Maho15a, Riec15b}. They also flag a starting point from which to understand
the statistical and structural properties of quantum-state time series. That
step will provide tools necessary not only for furthering our understanding of
fundamental quantum dynamics, but also grasping the operational meaning of
their informational properties in the context of quantum computation.

\section*{Acknowledgments}
\label{sec:acknowledgments}

We thank Fabio Anza, David Gier, Alex Jurgens, Sam Loomis, and participants of
the Telluride Science Research Center Information Engines Workshops for helpful
discussions. JPC acknowledges the kind hospitality of the Telluride Science
Research Center, Santa Fe Institute, Institute for Advanced Study at the
University of Amsterdam, and California Institute of Technology for their
hospitality during visits. This material is based upon work supported by, or in
part by, Grant Nos. FQXi-RFP-IPW-1902 and FQXi-RFP-1809 from the Foundational
Questions Institute and Fetzer Franklin Fund (a donor-advised fund of Silicon
Valley Community Foundation) and grants W911NF-18-1-0028 and W911NF-21-1-0048
from the U.S. Army Research Laboratory and the U.S. Army Research Office.

\section*{Conflicts of Interests}

The authors declare that they have no known competing financial interests or
personal relationships that could have appeared to influence the work reported
here.

\section*{Data Availability}

The datasets generated during and/or analyzed during the current study are available from the corresponding author on reasonable request.

\appendix

\section{Stochastic Processes}
\label{app:processes}

The Appendices provide a summary of and an introduction to the classical
theory that is the foundation for the main development. They cover basic
notation and definitions, as well as hidden Markov chains and their properties
and uses that are most relevant.

A classical stochastic process $\bf{X}$ is a bi-infinite series of indexed
observables produced by a system and is defined by the probability measure over
the random variables corresponding to the observables: $\Prob(\ldots X_{t-1},
X_t, X_{t+1}\ldots = \ldots x_{t-1}, x_t, x_{t+1} \ldots)$. In this, the random
variables corresponding to the observables are denoted with capital letters
$\ldots X_{t-2}, X_{t-1}, X_t, X_{t+1}, X_{t+2} \ldots$ and their realizations
are denoted by lowercase letters $\ldots x_{t-2}, x_{t-1}, x_t, x_{t+1},
x_{t+2} \ldots$, with the $x_t$ values drawn from a discrete alphabet $\Abet$.
The label $t$ in the indexing is chosen to evoke the traditional time-indexing
of stochastic processes in which the random variables are sequential
measurements of a physical system. Random variable blocks are denoted $X_{t:
t+l} = X_t, X_{t+1}, \ldots X_{t+l-1}$, with the left index inclusive and the
right exclusive. 

For present purposes, we concentrate on \emph{stationary} stochastic processes,
in which the joint distribution for a block of length $l$ is time (or
index)-translation invariant:
\begin{align*}
\Prob(X_{t:t+l} = x_{t:t+l}) = \Prob(X_{0:l} = x_{0:l}) 
  ~,
\end{align*}
for all $t$ and $l$.
A wide class of stationary stochastic processes can be modeled with Hidden
Markov Chains (HMCs) \cite{Rabi86a, Rabi89a, Crut92c}. Appendix \ref{app:hmm}
describes these in detail, reviewing how they facilitate calculating
various metrics for the stochastic processes they generate.

\section{Process Markovity}
\label{app:Markovity}

Given that the present setting considers both classical stochastic process
theory and quantum phenomena, it is helpful to comment on the concept of
Markovianity.

{\Def{A \emph{Markov process} or \emph{Markov chain} of order $\MOrder$
is a stationary stochastic process $\bf{X}$ in which the probability distribution satisfies the following:
\begin{align}
\Prob(& X_t = x_t| X_{-\infty : t} = x_{-\infty:t}) \nonumber \\
  & = \Prob(X_t = x_t| X_{t-\MOrder : t} = x_{t-\MOrder:t}) \nonumber \\
  & = \Prob(X_t=x_t| X_{t-1} = x_{t-1}\ldots X_{t-R} = x_{t-R})
  ,
\label{eqn:markov}
\end{align}
for all $t \in \mathbb{Z}$ and $R \in \mathbb{N}$.
}}

That is, the probability distribution of a particular random variable
conditioned on the past depends only on the value of the previous $\MOrder$
random variables. 

From this definition, we emphasize the following: 
\begin{itemize}
\setlength{\topsep}{-1pt}
\setlength{\itemsep}{-1pt}
\setlength{\parsep}{-1pt}
\item Memoryless or independent identically distributed (\emph{i.i.d.})
	processes are stochastic processes with $\MOrder =0$. We also refer to
	them as Markov processes of order $0$. 
\item Stochastic processes with $1 \leq \MOrder < \infty$ are memoryful. We
	refer to them as Markov processes of order $\MOrder$. 
\item Memoryful stochastic processes that do not satisfy the Markov condition
	in Eq. (\ref{eqn:markov}) for finite $\MOrder$, but require $\MOrder =
	\infty$ are infinite-order Markov processes. The latter is a surprisingly
	common property \cite{Jame10a}. They are the best candidates for the
	descriptor ``non-Markov processes''.
\end{itemize}
 
These points are necessary as there is contradictory use of the terminology in
the quantum non-Markovianity literature. In particular, there is extensive use
of the descriptor ``memoryless'' to refer to Markov processes with Markov
order $\MOrder =1$ and ``non-Markovian'' to refer to processes with finite
Markov order $R > 1$ \cite{Riva14a, Li18a, Poll18a, Tara19b, Milz20b}. While
those use cases are in dynamical settings distinct from the one we address
here it is pertinent to clarify our nomenclature.

To avoid confusion, the nomenclature above is directly derived from the
definition of Markov property, consistent with A. A. Markov's original
motivations to study ``complex chains'', his phrase for memoryful stochastic
processes \cite{Bash04a,Mark06a,Mark13a,Mark13b}. Memoryless is $\MOrder = 0$,
Markov is $1 \leq \MOrder < \infty$, and infinite Markov or non-Markov is
$\MOrder = \infty$. Reference \cite{Ara14a}'s introduction reviews this and the
early history of probabilistic chains.

\section{Hidden Markov Chains}
\label{app:hmm}

{\Def{A \emph{hidden Markov chain} (HMC) is a tuple ($\SSet$, $\Abet$, $\{T^x\}$) that consists of: 
\begin{itemize}
\setlength{\topsep}{-1pt}
\setlength{\itemsep}{-1pt}
\setlength{\parsep}{-1pt}
\item Set $\SSet$ of hidden states $\cs \in \SSet$.  
\item Discrete alphabet $\Abet$: a set of symbols that the HMC emits on
	state-to-state transitions at each time step. 
\item $\{T^{(x)}\} , ~x \in \Abet$ is the set of symbol-labeled transition
	matrices such that $T^{(x)}_{\cs\cs'} = \Prob(x, \cs'|\cs)$ with
	$\cs,\cs' \in \SSet$. 
\end{itemize}}}

The tuple directly defines the dynamic over the hidden states, which is itself
Markovian (order $R = 1$). That Markov chain's transition matrix is given by $T
= \sum_{x \in \Abet} T^x$. In turn, $T$ defines the stationary state
distribution $\pi$ over hidden states, such that $\pi \cdot T = \pi$. That is,
$\pi$ is a row vector such that $\pi_\cs = \Pr(\cs)$ with $\cs \in \SSet$.

See, for example, Fig. \ref{fig:ex12}, where $\SSet = \{A, B\}$ and $\Abet
=\{0,1\}$. 

An HMC then is a model for a stochastic process consisting of the set of emitted
symbol sequences and their associated probabilities. It is important to note
here that, even if the dynamic over hidden states is Markovian, HMCs generate
a more general class of stochastic processes that includes non-Markovian
processes. While the hidden states give an explicit mechanism for producing a
stochastic process, the stochastic process itself is defined only over the set
of symbols $x \in \Abet$. 

HMCs are useful in that they specify a finite mechanistic procedure to produce
the correct probabilities for a stochastic process. There is an infinite family
of distinct HMCs that model a given process. These are called a process'
\emph{presentations}. For each process, though, there is a unique presentation
with specific properties that, beyond merely generating the correct
probabilities, captures the minimal conditional distributions needed for
optimal prediction. This presentation is called a process' \eM, and it allows
for efficient exact computation of a process' randomness and structure metrics.
Appendix \ref{app:CMech} summarizes what an \eM is and why it is such an
important tool when modeling a stochastic process. 

In point of fact, the development here works with \emph{edge-labeled} HMCs.
There are also \emph{state-labeled} HMCs that emit symbols on entering a state.
Both presentation classes generate the same class of stochastic processes. They
are equivalent in this sense and they can be directly interconverted.
Edge-labeled presentations, though, do offer computational advantages when
calculating informational properties and in interpreting the functionality of
their operation.

\section{Computational Mechanics}
\label{app:CMech}

This section briefly reviews the main results of \emph{computational mechanics}
\cite{Crut88a, Shal98a, Crut12a}.

For ease of notation, we refer here to the past sequences of a process as
$\overleftarrow{X} = X_{-\infty:t}$ and the future sequences as
$\overrightarrow{X} = X_{t:\infty}$. For finite futures of length $\ell$ we use
$X_{t:t+\ell} = \overrightarrow{X}^\ell$.

As addressed in the main text, the hidden states in a unifilar presentation
must satisfy the condition that, given an observed past sequence
$\overleftarrow{x}$, all the allowed hidden states induced by that observation
must have the same distribution of futures
$\Prob(\overleftarrow{X}|\overleftarrow{x})$. If every observed past induces a
unique allowed state in a unifilar presentation, we call that a \emph{causal
state}.

{\Def{\emph{Causal states} are the equivalence classes of pasts determined by
the equivalence relation $\sim_\epsilon$. The latter defines two infinite past
sequences $\overleftarrow{x}$ and $\overleftarrow{x}'$ as
equivalent---$\overleftarrow{x} \sim_\epsilon \overleftarrow{x}'$---if and only
if they have the same conditional distribution of futures:
\begin{align*}
& \epsilon(\overleftarrow{x}) = \big\{\overleftarrow{x}'| \\
  & \quad \Prob(\overrightarrow{X}^\ell = \overrightarrow{x}^\ell |
	\overleftarrow{X} =\overleftarrow{x})
  =
 \Prob(\overrightarrow{X}^\ell = \overrightarrow{x}^\ell |
 	\overleftarrow{X} =\overleftarrow{x}')
	\big\}
  ~,
\end{align*}
where $\overrightarrow{x}^\ell \in \overrightarrow{X}^\ell$,
$\overleftarrow{x}' \in \overleftarrow{X}$, and $\ell \in \mathbb{Z}^+$. We
denote a causal state random variable by $\CausalState$, a particular causal
state of an HMC by $\causalstate$, and the set of causal states by
$\CausalStateSet$.
}}

As Ref. \cite{Shal98a} details, a given causal state and the next observed
symbol of a process determine a unique next causal state. A given causal state
$\causalstate$ also provides a well-defined conditional probability
$\Pr(\overrightarrow{X}|\causalstate)$ for all possible future sequences
$\overrightarrow{X}$. These two facts together mean there is a well-defined set
of labeled transition matrices $\{T^{(x)}\}$ that describe the probabilities of
transition between causal states given an observed symbol $x$. 

{\Def{The causal state set $\CausalStateSet$, together with the labeled
transition matrices $\{T^{(x)}: x \in \MeasAlphabet \}$ define a process'
\emph{\eM}. 
}}

For any given stochastic process, its \eM is unique. It is also the process
presentation with maximally accurate prediction of minimal statistical
complexity. This makes the \eM a natural canonical HMC for a process. 

\section{Mixed-State Presentations}
\label{app:msa}

The following introduces the \emph{Mixed State Algorithm} (MSA) that converts
a nonunifilar presentation to a unifilar presentation.

Assume that an observer has an HMC presentation $M$ for a process $\Process$
that emits symbols $x \in \MeasAlphabet$. Before making any observations, it
has probabilistic knowledge of the current state $\mxst_0 = \Prob(\SSet)$. We
call this a \emph{state of knowledge} or \emph{belief distribution}.
Typically, the best guess for an observer prior to observing any output of the
system is $\mxst_0 = \pi$. 

Once $M$ generates a word $w = x_0 x_1 \ldots x_\ell$ the observer's \emph{state of knowledge} of $M$'s current state can be updated to $\mxst(w)$, 
that is: 
\begin{align}
\mxst(w)_\causalstate & \equiv \Prob(\CausalState_\ell = \causalstate |
  \MS{0}{\ell}=w, \CausalState_0 \sim \pi)
  ~.
\label{eq:MixedState}
\end{align}

The collection of possible \emph{states of knowledge} $\mxst(w)$ forms $M$'s set $\MxSSet$ of \emph{mixed states}:
\begin{align*}
\MxSSet = \{ \mxst(w): w \in \MeasAlphabet^+, \Pr(w) > 0 \}
  ~,
\end{align*}
where $\MeasAlphabet^+$ is the set of all words with positive length. 

There is also the mixed-state measure $\MxSMeasure(\mxst)$---the probability
of being in a particular mixed state:
\begin{align*}
\Pr(\mxst(w)) & = \Prob(\CausalState_\ell |
  \MS{0}{\ell}=w, \CausalState_0 \sim \pi) \Pr(w)
  ~.
\end{align*}

From this follows the probability of transitioning from $\mxst(w)$ to
$\mxst(w\msym)$ on observing symbol $\msym$:
\begin{align*}
\Pr(\mxst(w\msym) | \mxst(w))
  & = \Pr(\msym|\CausalState_\ell \sim \mxst(w))
  ~.
\end{align*}
This defines the mixed-state dynamic $\MxSDyn$, in terms of the original
process not in terms of an HMC presentation of the latter.

Given any presentation $M$ of a process, we can calculate a new presentation
for the process with important properties as follows. The probability of generating symbol $\msym$ when in mixed state $\mxst$ is:
\begin{align}
\Pr(\msym |\mxst) = \mxst \cdot T^{(\msym)} \cdot \One 
  ~,
\label{eq:SymbolFromMixedState}
\end{align}
where $\One$ is a column vector of ones. Upon seeing symbol $\msym$, the
current mixed state $\mxst_t$ is updated:
\begin{align}
\mxst_{t+1}(x)= \frac{\mxst_t \cdot T^{(\msym)}} {\Prob(\msym |\mxst)}
  ~.
\label{eq:MxStUpdate}
\end{align}

Thus, given an HMC presentation we calculate the mixed state of Eq.
(\ref{eq:MixedState}) via:
\begin{align*}
\mxst(w) = \frac{\pi \cdot T^{(w)}}{\pi \cdot T^{(w)} \cdot \One}
  ~.
\end{align*}
And the mixed-state transition dynamic is then:
\begin{align*}
\Pr(\mxst_{t+1},\msym|\mxst_t) & = \Pr(\msym|\mxst_t) \\
   & = \mxst_t \cdot T^{(\msym)} \cdot \One
   ~,
\end{align*}
since Eq. (\ref{eq:MxStUpdate}) says that, by construction, the MSP is
unifilar. That is, the next mixed state is a function of the previous and the
emitted (observed) symbol. 

Together the mixed states and their dynamic give the HMC's \emph{mixed-state
presentation} (MSP) $\MSP = \{\MxSSet, \MxSDyn \}$, a unifilar presentation
for the process generated by presentation $M$.

\section{Measurement Angle Dependence}
\label{app:mad}

Two animations illustrate the measurement angle dependence of the MSP; see:\\
\url{https://csc.ucdavis.edu/~cmg/compmech/pubs/qdic.htm}.

The first animation shows shows how the mixed state presentation and $h_\mu$
vary as a function of parameter $\theta$. The second animation shows plots like
that in Fig. \ref{fig:rip_hmus}, while sweeping $\phi$ from $0$ to $2\pi$.


\begin{thebibliography}{10}

\bibitem{Schi11a}
P.~Schindler, J.T. Barreiro, T.~Monz, V.~Nebendahl, D.~Nigg, M.~Chwalla,
  M.~Hennrich, and R.~Blatt.
\newblock Experimental repetitive quantum error correction.
\newblock {\em Science}, 332(6033), 2011.

\bibitem{Nigg14a}
D.~Nigg, M.~M{\"u}ller, E.~A. Martinez, P.~Schindler, M.~Hennrich, T.~Monz,
  M.~A. Martin-Delgado, and R.~Blatt.
\newblock Quantum computations on a topologically encoded qubit.
\newblock {\em Science}, 345(6194):302--305, 2014.

\bibitem{Ofek16a}
N.~Ofek, A.~Petrenko, R.~Heeres, P.~Reinhold, Z.~Leghtas, B.~Vlastakis, Y.~Liu,
  L.~Frunzio, S.~M. Girvin, L.~Jiang, M.~Mirrahimi, M.~Devoret, and
  R.~Shoelkopf.
\newblock Extending the lifetime of a quantum bit with error correction in
  superconducting qubits.
\newblock {\em Nature}, 536:441--445, 2016.

\bibitem{Saro20a}
M.~Sarovar, T.~Proctor, K.~Rudinger, K.~Young, E.~Nielsen, and R.~Blume-Kohout.
\newblock Detecting crosstalk errors in quantum information processors.
\newblock {\em {Quantum}}, 4:321, 2020.

\bibitem{Harp20a}
R.~Harper, S.~T. Flammia, and J.~J. Wallman.
\newblock Efficient learning of quantum noise.
\newblock {\em Nature Phys.}, 16(1184-1188), 2020.

\bibitem{Riva14a}
A.~Rivas, S.~F. Huelga, and M.~B. Plenio.
\newblock Quantum {non-Markovianity}: characterization, quantification and
  detection.
\newblock {\em Rep. Prog. Phys.}, 77(094001), 2014.

\bibitem{deVe17a}
I.~de~Vega and D.~Alonso.
\newblock Dynamics of {non-Markovian} open quantum systems.
\newblock {\em Rev. Mod. Phys.}, 89:015001, Jan 2017.

\bibitem{Whit20a}
G.~A.~L. White, C.D. Hill, F.~A. Pollock, L.~C.~L. Hollenberg, and K.~Modi.
\newblock Demonstration of {non-Markovian} process characterization and control
  on a quantum processor.
\newblock {\em Nature Commun.}, 11(6301), 2020.

\bibitem{Vene20a}
A.~E. Venegas-Li, A.~M. Jurgens, and J.~P. Crutchfield.
\newblock Measurement-induced randomness and structure in controlled qubit
  processes.
\newblock {\em Phys. Rev. E}, 102:040102, 2020.

\bibitem{Jurg20b}
A.~Jurgens and J.~P. Crutchfield.
\newblock Shannon entropy rate of hidden {Markov} processes.
\newblock {\em J. Statistical Physics}, 183(32):1--18, 2020.

\bibitem{Jurg20c}
A.~Jurgens and J.~P. Crutchfield.
\newblock Divergent predictive states: The statistical complexity dimension of
  stationary, ergodic hidden {Markov} processes.
\newblock {\em Chaos}, 31(8):0050460, 2021.

\bibitem{Jurg20e}
A.~Jurgens and J.~P. Crutchfield.
\newblock Ambiguity rate of hidden {Markov} processes.
\newblock {\em Phys. Rev. E}, 104:064107, 2021.

\bibitem{Crut88a}
J.~P. Crutchfield and K.~Young.
\newblock Inferring statistical complexity.
\newblock {\em Phys. Rev. Let.}, 63:105--108, 1989.

\bibitem{Shal98a}
C.~R. Shalizi and J.~P. Crutchfield.
\newblock Computational mechanics: Pattern and prediction, structure and
  simplicity.
\newblock {\em J. Stat. Phys.}, 104:817--879, 2001.

\bibitem{Crut12a}
J.~P. Crutchfield.
\newblock Between order and chaos.
\newblock {\em Nature Physics}, 8(January):17--24, 2012.

\bibitem{Parr15a}
J.~Parrondo, J.~Horowitz, and T.~Sagawa.
\newblock Thermodynamics of information.
\newblock {\em Nature Physics}, 11:131--139, 2015.

\bibitem{Cont19a}
T.~Conte, E.~DeBenedictis, N.~Ganesh, T.~Hylton, J.~Paul Strachan, R.~S.
  Williams, A.~Alemi, L.~Altenberg, G.~E. Crooks, J.~P. Crutchfield, L.~del
  Rio, J.~Deutsch, M.~R. DeWeese, K.~Douglas, M.~Esposito, M.~P. Frank, R.~Fry,
  P.~Harsha, M.~D. Hill, C.~Kello, J.~Krichmar, S.~Kumar, S.-C. Liu, S.~Lloyd,
  M.~Marsili, I.~Nemenman, A.~Nugent, N.~Packard, D.~Randall, P.~Sadowski,
  N.~Santhanam, R.~Shaw, A.~Z. Stieg, E.~Stopnitzky, C.~Teuscher, C.~Watkins,
  D.~Wolpert, J.~J. Yang, and Y.~Yufik.
\newblock Thermodynamic computing.
\newblock {\em CoRR}, abs/1911.01968, 2019.

\bibitem{Gard04a}
C.~Gardiner and P.~Zoller.
\newblock {\em Quantum Noise: A Handbook of Markovian and Non-Markovian Quantum
  Stochastic Methods with Applications to Quantum Optics}.
\newblock Springer, 2004.

\bibitem{Cler10a}
A.~A. Clerk, M.~H. Devoret, S.~M. Girvin, Florian Marquardt, and R.~J.
  Schoelkopf.
\newblock Introduction to quantum noise, measurement, and amplification.
\newblock {\em Rev. Mod. Phys.}, 82:1155--1208, Apr 2010.

\bibitem{Breu07a}
H.-P. Breuer and F.~Petruccione.
\newblock {\em The Theory of Open Quantum Systems}.
\newblock Oxford University Press, Oxford, United Kingdom, 2007.

\bibitem{Riva12a}
A.~Rivas and S.~F. Huelga.
\newblock {\em Open Quantum Systems: An Introduction}.
\newblock Springer, Heidelberg, Germany, 2012.

\bibitem{Breu16a}
H.-P. Breuer, E.-M. Laine, J.~Piilo, and B.~Vacchini.
\newblock {Non-Markovian} dynamics in open quantum systems.
\newblock {\em Rev. Mod. Phys.}, 88:021002, Apr 2016.

\bibitem{Li18a}
L.~Li, M.~J.~W. Hall, and H.~Wiseman.
\newblock Concepts of quantum {non-Markovianity}: A hierarchy.
\newblock {\em Phys. Rep.}, 759:1--51, 2018.

\bibitem{Mark13a}
A.~A. Markov.
\newblock Primer statisticheskogo issledovaniya nad tekstom {``Evgeniya
  Onegina''}, illyustriruyuschij svyaz' ispytanij v cep'.
\newblock {\em Izv. Akad. Nauk, SPb}, 93:153--162, 1913.

\bibitem{Mark13b}
A.~A. Markov.
\newblock {\em Ischislenie veroyatnostej}.
\newblock Spb, 1900; 2-e izd., spb, 1908 edition, 1913.
\newblock Translated into German, Wahrscheinlichkeits-Rechnung, Teubner,
  Leipzig-Berlin, 1912; 3-e izd., SPb, 1913; 4-e izd., Moskva, 1924.

\bibitem{Milz20a}
S.~Milz and K.~Modi.
\newblock Quantum stochastic processes and quantum {non-Markovian} phenomena,
  2020.

\bibitem{Poll18a}
F.~A. Pollock, C.~Rodr\'{\i}guez-Rosario, T.~Frauenheim, M.~Paternostro, and
  K.~Modi.
\newblock {Non-Markovian} quantum processes: Complete framework and efficient
  characterization.
\newblock {\em Phys. Rev. A}, 97:012127, 2018.

\bibitem{Poll18b}
F.~A. Pollock, C.~Rodr\'{\i}guez-Rosario, T.~Frauenheim, M.~Paternostro, and
  K.~Modi.
\newblock Operational {Markov} condition for quantum processes.
\newblock {\em Phys. Rev. Lett.}, 120:040405, 2018.

\bibitem{Tara19a}
P.~Taranto, S.~Milz, F.~A. Pollock, and K.~Modi.
\newblock Structure of quantum stochastic processes with finite {Markov} order.
\newblock {\em Phys. Rev. A}, 99:042108, 2019.

\bibitem{Efro16a}
A.~Efros and D.~Nesbitt.
\newblock Origin and control of blinking in quantum dots.
\newblock {\em Nature Nanotech}, 11:661--671, 2016.

\bibitem{Crut01a}
J.~P. Crutchfield and D.~P. Feldman.
\newblock Regularities unseen, randomness observed: Levels of entropy
  convergence.
\newblock {\em CHAOS}, 13(1):25--54, 2003.

\bibitem{Rabi86a}
L.~R. Rabiner and B.~H. Juang.
\newblock An introduction to hidden {Markov} models.
\newblock {\em IEEE ASSP Magazine}, January:4--16, 1986.

\bibitem{Rabi89a}
L.~R. Rabiner.
\newblock A tutorial on hidden {Markov} models and selected applications.
\newblock {\em IEEE Proc.}, 77:257, 1989.

\bibitem{Bech15a}
J.~Bechhoefer.
\newblock Hidden {Markov} models for stochastic thermodynamics.
\newblock {\em New. J. Phys.}, 17:075003, 2015.

\bibitem{Riec13a}
P.~Riechers and J.~P. Crutchfield.
\newblock Spectral simplicity of apparent complexity, {Part I}: {The}
  nondiagonalizable metadynamics of prediction.
\newblock {\em Chaos}, 28:033115, 2018.

\bibitem{Riec17a}
P.~Riechers and J.~P. Crutchfield.
\newblock Spectral simplicity of apparent complexity, {Part II}: {Exact}
  complexities and complexity spectra.
\newblock {\em Chaos}, 28:033116, 2018.

\bibitem{Moor97a}
C.~Moore and J.~P. Crutchfield.
\newblock Quantum automata and quantum grammars.
\newblock {\em Theoret. Comp. Sci.}, 237:1-2:275--306, 2000.

\bibitem{Gier21a}
D.~Gier.
\newblock Adaptive measurement protocols.
\newblock {\em in preparation}, 2023.

\bibitem{Blac57b}
D.~Blackwell.
\newblock The entropy of functions of finite-state {Markov} chains.
\newblock volume~28, pages 13--20, Publishing House of the Czechoslovak Academy
  of Sciences, Prague, 1957.
\newblock Held at Liblice near Prague from November 28 to 30, 1956.

\bibitem{Crut08a}
J.~P. Crutchfield, C.~J. Ellison, and J.~R. Mahoney.
\newblock Time's barbed arrow: {Irreversibility}, crypticity, and stored
  information.
\newblock {\em Phys. Rev. Lett.}, 103(9):094101, 2009.

\bibitem{Crut08b}
C.~J. Ellison, J.~R. Mahoney, and J.~P. Crutchfield.
\newblock Prediction, retrodiction, and the amount of information stored in the
  present.
\newblock {\em J. Stat. Phys.}, 136(6):1005--1034, 2009.

\bibitem{Crut92c}
J.~P. Crutchfield.
\newblock The calculi of emergence: Computation, dynamics, and induction.
\newblock {\em Physica D}, 75:11--54, 1994.

\bibitem{Marz17a}
S.~E. Marzen and J.~P. Crutchfield.
\newblock Nearly maximally predictive features and their dimensions.
\newblock {\em Phys. Rev. E}, 95(5):051301(R), 2017.

\bibitem{Niel11a}
M.~A. Nielsen and I.~L. Chuang.
\newblock {\em Quantum Computation and Quantum Information}.
\newblock Cambridge University Press, Cambridge, United Kingdom, tenth
  anniversary edition, 2011.

\bibitem{Shan48a}
C.~E. Shannon.
\newblock A mathematical theory of communication.
\newblock {\em Bell Sys. Tech. J.}, 27:379--423, 623--656, 1948.

\bibitem{Cent94a}
P.~M. Centore and E.~R. Vrscay.
\newblock Continuity of attractors and invariant measures for iterated function
  systems.
\newblock {\em Canadian Math. Bull.}, 37(3):315?329, 1994.

\bibitem{Mend98a}
F.~Mendivil.
\newblock A generalization of ifs with probabilities to infinitely many maps.
\newblock {\em Rocky Mountain J. Math.}, 28(3):1043--1051, 1998.

\bibitem{Kloe20a}
B.~Kloeckner.
\newblock {Optimal transportation and stationary measures for Iterated Function
  Systems}.
\newblock {\em {Mathematical Proceedings}}, 2021.
\newblock https://hal.archives-ouvertes.fr/hal-02276750.

\bibitem{Ivan87a}
I.~D. Ivanovic.
\newblock How to differentiate between non-orthogonal states.
\newblock {\em Phys. Let. A}, 123(6):257--259, 1987.

\bibitem{Diek88a}
D.~Dieks.
\newblock Overlap and distinguishability of quantum states.
\newblock {\em Phys. Let. A}, 126(5):303--306, 1988.

\bibitem{Pere88a}
A.~Peres.
\newblock How to differentiate between non-orthogonal states.
\newblock {\em Phys. Let. A}, 128(1):19, 1988.

\bibitem{Feld08a}
D.~P. Feldman, C.~S. McTague, and J.~P. Crutchfield.
\newblock The organization of intrinsic computation: {Complexity}-entropy
  diagrams and the diversity of natural information processing.
\newblock {\em CHAOS}, 18(4):043106, 2008.

\bibitem{Sina59}
Ja.~G. Sinai.
\newblock On the notion of entropy of a dynamical system.
\newblock {\em Dokl. Akad. Nauk. SSSR}, 124:768, 1959.

\bibitem{Cove91a}
T.~M. Cover and J.~A. Thomas.
\newblock {\em Elements of Information Theory}.
\newblock Wiley-Interscience, New York, 1991.

\bibitem{Falk95a}
M.~Falkhausen, H.~Reininger, and D.~Wolf.
\newblock Calculation of distance measures between hidden {Markov} models.
\newblock In {\em EUROSPEECH}, 1995.

\bibitem{Sahr10a}
S.~M.~E. Sahraeian and B.-J. Yoon.
\newblock A novel low-complexity {HMM} similarity measure.
\newblock {\em IEEE Signal Proc. Let.}, 18(2):87--90, 2011.

\bibitem{Zeng10a}
J.~Zeng, J.~Duana, and C.~Wu.
\newblock A new distance measure for hidden {Markov} models.
\newblock {\em Expert Systems with Applications}, 37(2):1550--1555, 2010.

\bibitem{Riec18a}
P.~M. Riechers and J.P. Crutchfield.
\newblock Beyond the spectral theorem: Decomposing arbitrary functions of
  non-diagonalizable operators.
\newblock {\em A.I.P Advances}, 8:065305, 2018.

\bibitem{Cros08a}
G.~M. Crosswhite and D.~Bacon.
\newblock Finite automata for caching in matrix product algorithms.
\newblock {\em Phys. Rev. A}, 78(012356), 2008.

\bibitem{Scho05a}
C.~Sch{\"o}n, E.~Solano, F.~Verstraete, J.~I. Cirac, and M.~M. Wolf.
\newblock Sequential {Generation} of {Entangled} {Multiqubit} {States}.
\newblock {\em Phys. Rev. Let.}, 95(11):110503, 2005.

\bibitem{Scho07a}
C.~Sch{\"o}n, K.~Hammerer, M.~M. Wolf, J.~I. Cirac, and E.~Solano.
\newblock Sequential generation of matrix-product states in cavity {QED}.
\newblock {\em Phys. Rev. A}, 75(3):032311, 2007.

\bibitem{Marz17b}
S.~Marzen and J.~P. Crutchfield.
\newblock Structure and randomness of continuous-time discrete-event processes.
\newblock {\em J. Stat. Physics}, 169(2):303--315, 2017.

\bibitem{Gu12a}
M.~Gu, K.~Wiesner, E.~Rieper, and V.~Vedral.
\newblock Quantum mechanics can reduce the complexity of classical models.
\newblock {\em Nature Comm.}, 3(762):1--5, 2012.

\bibitem{Maho15a}
J.~R. Mahoney, C.~Aghamohammadi, and J.~P. Crutchfield.
\newblock Occam's quantum strop: {Synchronizing} and compressing classical
  cryptic processes via a quantum channel.
\newblock {\em Scientific Reports}, 6:20495, 2016.

\bibitem{Riec15b}
P.~M. Riechers, J.~R. Mahoney, C.~Aghamohammadi, and J.~P. Crutchfield.
\newblock Minimized state-complexity of quantum-encoded cryptic processes.
\newblock {\em Phys. Rev. A}, 93(5):052317, 2016.

\bibitem{Jame10a}
R.~G. James, J.~R. Mahoney, C.~J. Ellison, and J.~P. Crutchfield.
\newblock Many roads to synchrony: Natural time scales and their algorithms.
\newblock {\em Phys. Rev. E}, 89:042135, 2014.

\bibitem{Tara19b}
P.~Taranto, F.~A. Pollock, S.~Milz, M.~Tomamichel, and K.~Modi.
\newblock Quantum {Markov} order.
\newblock {\em Phys. Rev. Lett.}, 122:140401, 2019.

\bibitem{Milz20b}
S.~Milz, D.~Egloff, P.~Taranto, T.~Theurer, M.~B. Plenio, A.~Smirne, and S.~F.
  Huelga.
\newblock When is a {Non-Markovian} quantum process classical?
\newblock {\em Phys. Rev. X}, 10:041049, 2020.

\bibitem{Bash04a}
G.~P. Basharin, A.~N. Langville, and V.~A. Naumov.
\newblock The life and work of {A. A. Markov}.
\newblock {\em Linear Algebra and its Applications}, 386:3--26, 2004.

\bibitem{Mark06a}
A.~A. Markov.
\newblock An example of statistical investigation of the text ``{Eugene
  Onegin}'' concerning the connection of samples in chains.
\newblock {\em Science in Context}, 19:591--600, 2006.

\bibitem{Ara14a}
P.~M. Ara, R.~G. James, and J.~P. Crutchfield.
\newblock The elusive present: {Hidden} past and future dependence and why we
  build models.
\newblock {\em Phys. Rev. E}, 93(2):022143, 2016.

\end{thebibliography}
\end{document}